\documentclass[aps,prd,tightenlines,twocolumn,amsmath,amssymb,showpacs]{revtex4}%
\usepackage{graphicx}
\usepackage{epsfig}
\usepackage{subfigure}
\usepackage{graphics,color}
\usepackage[none]{hyphenat}
\usepackage{hyperref}
\usepackage{amsbsy}
\usepackage{bm}
\newcommand{\be}{\begin{equation}}
\newcommand{\ee}{\end{equation}}
\newcommand{\bea}{\begin{eqnarray}}
\newcommand{\eea}{\end{eqnarray}}
\newcommand{\nn}{\nonumber}
\newcommand{\Tr}{{\mathrm{Tr}}}
\def \l{\left(}
\def \r{\right)}
\begin{document}
\title{Comparing symmetry restoration trends for meson masses and mixing angles in the QCD-like three quark flavor models}
\author{Vivek Kumar Tiwari}
\email {vivekkrt@gmail.com}
\affiliation{Department of Physics, University of Allahabad, Allahabad 211002, India.}
\date{\today}
\begin{abstract}
We are computing the modifications for the  scalar and pseudoscalar meson masses and mixing angles due to
the proper accounting of fermionic vacuum fluctuation in the framework of the generalized $2+1$ flavor quark meson
model and the Polyakov loop augmented quark meson model(PQM). The renormalized contribution  of the divergent fermionic
vacuum fluctuation at one loop level makes these models effective QCD-like models. It has 
been explicitly shown that analytical expressions for the  model parameters, meson masses, and mixing angles do not
depend on any arbitrary renormalization scale. We have investigated how the incorporation of fermionic vacuum fluctuation
in quark meson and PQM models qualitatively and quantitatively affects the convergence in the masses of the chiral partners
in pseudoscalar ($\pi$, $\eta$, $\eta'$, $K$) and scalar ($\sigma$, $a_0$, $f_0$, $\kappa$) meson nonets as the temperature
is varied on the reduced temperature scale. Comparison of present results in the quark meson model with vacuum term  and 
PQM model with vacuum term with the already existing calculations in the bare $2+1$ quark meson and PQM models, shows that
the restoration of chiral symmetry becomes smoother due to the influence of the fermionic vacuum term. We find that the melting 
of the strange condensate registers a significant increase in the presence of the fermionic vacuum term and its highest melting is 
found in the PQM model with vacuum term. The role of the $U_A(1)$ anomaly in determining the isoscalar masses and mixing angles for the pseudoscalar
($\eta$ and $\eta'$) and scalar ($\sigma$ and $f_0$) meson complex has also been significantly modified due to the fermionic vacuum
correction. In its influence, the interplay of chiral symmetry restoration and the setting up of the $U_A(1)$ restoration 
trends have also been shown to be significantly modified.
\end{abstract}
\maketitle
\section{Introduction}
\label{intr}
The strong interaction theory predicts that normal hadronic matter 
goes through a phase transition and produces a collective form of matter 
known as the Quark Gluon Plasma (QGP)  under the extreme conditions of 
high temperature and/or density when the individual hadrons dissolve 
into their quark and gluon constituents \cite{Shuryak,Rafelski,Svetitsky,Ortmanns:96ea,Muller,Rischke:03}.
Relativistic heavy ion collision experiments at RHIC (BNL), LHC (CERN) and 
the future CBM experiments at the FAIR facility (GSI-Darmstadt) aim to create
and study such a collective state of matter. Study of the different aspects of 
this phase transition, is a tough and challenging task because Quantum Chromodynamics (QCD)
which is the theory of strong interaction, becomes nonperturbative in the low energy limit.
However the QCD vacuum reveals itself through the process of spontaneous chiral symmetry
breaking and phenomenon of color confinement.
 
	In the zero quark mass limit, chiral condensate works as an order 
parameter for the spontaneous breakdown of the chiral symmetry in the 
low energy hadronic vacuum of the QCD. For the infinitely heavy quarks, 
in the pure gauge $SU_c(3)$ QCD, the $Z(3)$ (Center symmetry of the QCD 
color gauge group) symmetry, which is the symmetry of hadronic vacuum, 
gets spontaneously broken in the high temperature/density regime of QGP. 
Here the expectation value of the Wilson line (Polyakov loop) is related 
to the free energy of a static color charge, hence it serves as the order
parameter of the confinement-deconfinement phase transition 
\cite{Polyakov:78plb}. Even though the center symmetry is always broken 
with the inclusion of dynamical quarks in the system, one can regard the 
Polyakov loop as an approximate order parameter because it is a good 
indicator of the confinement-deconfinement transition 
\cite{Pisarski:00prd,Vkt:06}.
 
	The lattice QCD calculations
(see e.g.~\cite{Kaczmarek:02,Karsch:02,Fodor:03,Allton:05,Karsch:05,
Karsch:07ax,Aoki:06,FodorN,Fodor1,HotQCD,ChengLQCD,BazLQCD,WBLQCD,LQCDWB2,LQCDWBL,HotLQCDL}) give us important information and insights
regarding various aspects of the transition, like the restoration of chiral 
symmetry in QCD, order of the confinement-deconfinement phase transition, 
richness of the QCD phase structure and mapping of the phase diagram. Since 
lattice calculations are technically involved and various issues are not conclusively
settled within the lattice community, one resorts to the calculations within the
ambit of phenomenological models \cite{Wilczek,Hatsuda:98,Bielich:00prl,Rischke:00,Mocsy:01prc,
Herpay:05,Herpay:06,Herpay:07,Schaefer:07prd,Schaefer:09,Schaefer:08ax,
Bowman:2008kc,koch,Andersen,Fejos:8285,Fejos} developed in terms of effective degrees 
of freedom. These model investigations complement the lattice simulation studies and 
give much needed insight about the regions of phase diagram inaccessible 
to lattice simulations. Lot of current effective model building activity, 
is centered around combining  the features of spontaneous breakdown of both 
chiral symmetry as well as the center $Z(3)$ symmetry of QCD in one single 
model (see for example \cite{Schaefer:07,Braun,gupta,Schaefer:09ax,Schaefer:09wspax,H.mao09,
Herbst:2010rf, Pawl:Schaef, Marko:2010cd,kahara,Digal:01,Ratti:06,
Ratti:07,Hansen:07,Ratti:07npa,Tamal:06,Sasaki:07,Hell:08,Abuki:08,Ciminale:07,
Fu:07,Kfuku:04plb,Fukushima:08d77,Fukushima:08d78,Fukushima:09,Contrera,
nonlocal,Odilon}). In these models chiral condensate and Polyakov loop are simultaneously 
coupled to the quark degrees of freedom. 

The behavior patterns of mesons and their properties in the hot and dense medium,
have been investigated in the several two and three flavor Nambu-Jona-Lasinio (NJL), 
Polyakov Nambu-Jona-Lasinio (PNJL) models (e.g.~\cite{Ratti:07prd,Costa:04,Costa:05,Costa:09,Contrera:NLM,Hiller}) and 
also in the $SU(2)$ version of linear sigma model (e.g.
\cite{Bielich:00prl,Mocsy:01prc,Schaefer:07prd}). Since the parity doubling
of mesons signals the restoration of chiral symmetry, these studies look for the
emergence of mass convergence patterns in the masses of the chiral partners
in pseudo scalar ($\pi$, $\eta$, $\eta'$, $K$) and scalar mesons
($\sigma$, $a_0$, $f_0$, $\kappa$). We know that the basic QCD Lagrangian has 
the global $SU_{R+L}(3) \times SU_{R-L}(3) \times U_A(1)$ symmetry. 
For the $SU(3)$  Linear Sigma Model, several explicit as well as spontaneous symmetry 
breaking patterns of $SU_{V}(3) \times SU_A(3)$, have been discussed by Lenaghan et. 
al. in Ref. \cite{Rischke:00}. Enlarging the Linear Sigma Model with the inclusion 
of quarks \cite{Schaefer:09} in the 2+1 flavor  breaking scenario, 
Schaefer et. al.  studied  the consequences of $SU(3)$ chiral symmetry 
restoration for scalar and pseudo scalar meson masses and mixing angles, 
in the presence as well as the absence of $U_A(1)$ axial symmetry, as 
the temperature is increased through the phase transition temperature. 
It was shown by 't Hooft \cite{tHooft:76prl} that the $U_A(1)$ axial symmetry
does not exist at the quantum level and the instanton effects explicitly break it
to $Z_{A}(N_f)$. Due to the $U_A(1)$ anomaly, the $\eta'$ meson
does not remain massless Goldstone boson in the chiral limit of zero quark masses
and it acquires a mass of about 1 GeV. This happens due to the flavor 
mixing, a phenomenon that lifts the degeneracy between the $\pi$ and 
$\eta'$ which otherwise would have been degenerate with $\pi$ in $U(3)$
even if the explicit chiral symmetry breaking is present.
There is large violation in Okubo-Zweig-Iizuka (OZI) rule for both pseudo scalar
and scalar mesons and ideal mixing is not achieved because of strong flavor mixing 
between non strange and strange flavor components of the mesons 
\cite{Costa:09}. Hence $U_A(1)$ restoration will have important observable 
effects on scalar and pseudo scalar meson masses as well as the mixing 
angles. 

The effect of Polyakov loop potential on the behavior of meson masses 
and mixing angles has been studied by Costa et. al. in 
the PNJL model \cite{Costa:09} and by Contrera et. al. in the nonlocal PNJL model \cite{Contrera:NLM}. 
Here in the NJL model based studies, mesons are generated by some prescription 
\cite{Costa:05} and the $\eta'$ is not a well defined quantity \cite{fukushima:01}. 
It becomes unbound soon after the temperature is raised from zero. 
In the 2+1 flavor quark meson linear sigma model investigations by Schaefer 
et al. \cite{Schaefer:09,Schaefer:08ax}, the mesons are the explicit degrees of 
freedom included in the Lagrangian from the very outset and 
the $U_A(1)$ breaking 't Hooft coupling term is constant.  
Recently, we investigated the influence of the Polyakov loop potential on the meson 
mass and mixing angle variations in the scalar and pseudo scalar sector, in the framework 
of generalized 2+1 flavor quark meson model enlarged with the inclusion of 
Polyakov loop \cite{gupta,Schaefer:09ax,Schaefer:09wspax,H.mao09}.

The chiral symmetry breaking mechanism in the Quark-Meson/Polyakov-Quark-Meson (QM/PQM) model is different 
from that of the NJL/PNJL model.
In the NJL/PNJL model, the fermionic vacuum fluctuation leads to the dynamical breaking of the chiral symmetry 
while in most of the QM/PQM model calculations, fermionic vacuum loop contribution to the grand potential has
frequently been neglected till recently \cite{Mocsy:01prc,Schaefer:07prd,Schaefer:09,Schaefer:08ax,kahara,Bowman:2008kc} because
here, the spontaneous breaking of chiral symmetry is generated by the mesonic potential itself.  
Recently, Skokov et. al.  incorporated the appropriately renormalized fermionic vacuum fluctuation 
\cite{Skokov:2010sf} in the thermodynamic potential of the two flavor QM model which becomes 
an effective QCD-like model because now it can reproduce the second order chiral phase transition
at $\mu=0$ as expected from the universality arguments \cite{Wilczek} for the two massless flavors of QCD.
The fermionic vacuum correction and its influence has also been 
investigated in earlier works \cite{Mizher:2010zb,Palhares:2008yq,Fraga:2009pi,Palhares:2010be}.
In a recent work \cite{Vivek:12}, we generalized the proper accounting of renormalized fermionic vacuum 
fluctuation in the two flavor PQM model to the non-zero chemical potentials and found that the position
of critical end point shifts to a significantly higher chemical potential in the $\mu$ and $T$ plane 
of the phase diagram. Very recently, Schaefer et. al. \cite{Schaef:12} estimated the size of critical 
region around the critical end point in a three flavor PQM model in the presence of the fermionic vacuum term.
Sandeep et al. also investigated the phase structure and made comparisons with lattice data 
in another recent 2+1 quark flavor study  with the effect of fermionic vacuum term \cite{Sandeep}. 
In a very recent work \cite{VKTR}, the present author explored and compared the details of criticality 
in the two flavor QM, PQM models in the presence and absence of fermionic vacuum correction.

In the present work, the author will explore how the proper accounting 
of fermionic vacuum correction in the QM and PQM models, 
qualitatively and quantitatively affects the convergence of the masses 
of chiral partners, when the parity doubling takes place as the 
temperature is increased through $T_c$ and the partial restoration of 
chiral symmetry is achieved. We will also be studying the effect of 
fermionic vacuum correction on the interplay of $SU_A(3)$ chiral symmetry and 
$U_A(1)$ symmetry restoration in the presence as well as absence of Polyakov loop
potential in QM model.
 
The arrangement of this paper is as follows. In Sec.\ref{sec:model}, we recapitulate the model formulation.
The grand potential in the mean field approach has been described in the Sec.
\ref{sec:potmf} where the subsection \ref{subsec:Vterm} explicitly explains the procedure for obtaining the scale independent 
expression of the effective potential after renormalizing the one loop fermionic vacuum
fluctuation. The numerical values of the model parameters are also given in this subsection while the mathematical
details for determining the renormalization scale independent parameters are given in the appendix A.
The final expressions of renormalization scale independent vacuum meson masses, are derived in 
the appendix B. The Sec.\ref{sec:mixang} gives the model formulae of meson masses and mixing angles 
in a finite temperature/density medium. In Sec.\ref{sec:eploop}, we will be discussing the numerical results
and plots for understanding and analyzing the effect of fermionic vacuum correction on the chiral symmetry restoration.  
Summary and conclusion is presented in the last Sec.\ref{sec:smry}.
\section{Model Formulation}
\label{sec:model}
We will be working in the generalized three flavor Quark Meson Chiral 
Linear Sigma Model which has been combined with the Polyakov loop 
potential
\cite{gupta,Schaefer:09wspax,H.mao09,Schaefer:09ax}. 
In this model, quarks coming in three flavor are coupled to the 
$SU_V(3) \times SU_A(3)$ symmetric mesonic fields together with 
spatially constant temporal gauge field represented by Polyakov loop 
potential. Polyakov loop field $\Phi$ is defined as the 
thermal expectation value of color trace of Wilson loop in temporal 
direction 
\be
\Phi = \frac{1}{N_c} \langle \Tr_c L(\vec{x})\rangle, \qquad \qquad  \Phi^* = \frac{1}{N_c} \langle \Tr_c L^{\dagger}(\vec{x}) \rangle
\ee
where $L(\vec{x})$ is a matrix in the fundamental representation of the 
$SU_c(3)$ color gauge group.
\be
\label{eq:Ploop}
L(\vec{x})=\mathcal{P}\mathrm{exp}\left[i\int_0^{\beta}d \tau
A_0(\vec{x},\tau)\right]
\ee
Here $\mathcal{P}$ is path ordering,  $A_0$ is the temporal vector 
field and $\beta = T^{-1}$ \cite{Polyakov:78plb}.
 
The model Lagrangian is written in terms of quarks, mesons, couplings 
and Polyakov loop potential ${\cal U} \left( \Phi, \Phi^*, T \right)$.

\be
\label{eq:Lag}
{\cal L}_{PQM} = {\cal L}_{QM} - {\cal U} \big( \Phi , \Phi^* , T \big) 
\ee
where the Lagrangian in Quark Meson Chiral Sigma model
\bea
\label{eq:Lqms}
{\cal L}_{QM} =  \bar{q_f} \big( i \gamma^{\mu}D_{\mu}  - g\; T_a\big( \sigma_a 
+ i\gamma_5 \pi_a\big) \big) q_f  + {\cal L}_{m} 
\eea
The coupling of quarks with the uniform temporal background gauge 
field is effected by the following replacement 
$D_{\mu} = \partial_{\mu} -i A_{\mu}$ 
and  $A_{\mu} = \delta_{\mu 0} A_0$ (Polyakov gauge), where 
$A_{\mu} = g_s A^{a}_{\mu} \lambda^{a}/2$ with vector potential $A^{a}_{\mu}$ for color gauge field. $g_s$ is the $SU_c(3)$ 
gauge coupling. $\lambda_a$ are Gell-Mann matrices in the color 
space, $a$ runs from $1 \cdots 8$. $q_f=(u,d,s)^T$ denotes the quarks 
coming in three flavors and three colors. $T_a$ represent 9 generators of $U(3)$ flavor symmetry with 
$T_a = \frac{\lambda_a}{2}$ and $a=0,1 \dots ~8$, here $\lambda_a$ are 
standard Gell-Mann matrices in flavor space with $\lambda_0=\sqrt{\frac{2}{3}}\ \bf 1$. $g$ is the flavor blind 
Yukawa coupling that couples the three flavor of quarks with nine 
mesons in the scalar ($\sigma_a, J^{P}=0^{+}$) and pseudo scalar 
($\pi_a, J^{P}=0^{-}$) sectors.

The quarks have no intrinsic mass but become massive after 
spontaneous chiral symmetry breaking because of non vanishing 
vacuum expectation value of the chiral condensate. The mesonic 
part of the Lagrangian has the following form
\bea  
\label{eq:Lagmes}
{\cal L}_{m} & = &\Tr \left( \partial_\mu M^\dagger \partial^\mu
    M \right)
  - m^2 \Tr ( M^\dagger M) -\lambda_1 \left[\Tr (M^\dagger M)\right]^2 \nn \\
  &&  - \lambda_2 \Tr\left(M^\dagger M\right)^2
  +c   \big[\text{det}(M) +\text{det} (M^\dagger) \big] \nn \\
  && + \Tr\left[H(M + M^\dagger)\right].
\eea
The chiral field $M$ is a $3 \times 3$ complex matrix comprising of 
the nine scalars $\sigma_a$ and the nine pseudo scalar $\pi_a$ mesons.
\be
\label{eq:Mfld}
M = T_a \xi_a= T_a(\sigma_a +i\pi_a)
\ee 
The generators follow $U(3)$ algebra $\left[T_a, T_b\right]  = if_{abc}T_c$ 
and $\left\lbrace T_a, T_b\right\rbrace  = d_{abc}T_c$ where 
$f_{abc}$ and $d_{abc}$ are standard antisymmetric and symmetric 
structure constants respectively with 
$f_{ab0}=0$ and $d_{ab0}=\sqrt{\frac{2}{3}} \ {\bf 1}\ \delta_{ab}$ 
and matrices are normalized as $\Tr(T_a T_b)=\frac{\delta_{ab}}{2}$.

The $SU_L(3) \times SU_R(3)$ chiral symmetry is explicitly broken by 
the explicit symmetry breaking term 
\be
H = T_a h_a
\ee
Here $H$ is a $3 \times 3$ matrix with nine external parameters. 
The $\xi$ field which denotes both the scalar as well as pseudo scalar mesons,
picks up the nonzero vacuum expectation value, $\bar{\xi}$ for the scalar mesons 
due to the spontaneous breakdown of the chiral symmetry while the pseudo scalar mesons
have zero vacuum expectation value. Since $\bar{\xi}$ must have the quantum numbers of the vacuum, 
explicit breakdown of the chiral symmetry is only possible with 
three nonzero parameters $h_0$, $h_3$ and $h_8$. We are neglecting 
isospin symmetry breaking hence we choose $h_0$, $h_8 \neq 0$.
This leads to the $2+1$ flavor symmetry breaking scenario with 
nonzero condensates $\bar{\sigma_0}$ and $\bar{\sigma_8}$. 

Apart from $h_0$ and $h_8$, the other parameters in the model are 
five in number. These are the squared tree-level mass of the meson 
fields $m^2$, quartic coupling constants $\lambda_1$ and $\lambda_2$, 
a Yukawa coupling $g$ and a cubic coupling constant $c$ which models 
the $U_A(1)$ axial anomaly of the QCD vacuum.
 
Since it is broken by the quantum effects, the $U_A(1)$ axial 
which otherwise is a symmetry of the classical Lagrangian, becomes
anomalous~\cite{Weinberg:75} and gives large mass to $\eta'$ meson 
($m_{\eta'} = 940$ MeV). In the absence of $U_A(1)$ anomaly, $\eta'$ 
meson would have been the ninth pseudo scalar Goldstone boson, 
resulting due to the spontaneous break down of the chiral $U_A(3)$ 
symmetry. The entire pseudo scalar nonet corresponding to the spontaneously 
broken $U_A(3)$, would consist of the three $\pi$, four $K$, $\eta$ 
and $\eta'$ mesons, which are the massless pure Goldstone
modes when $H = 0$ and they become pseudo Goldstone modes after 
acquiring finite mass due to nonzero $H$ in different symmetry 
breaking scenarios. The particles coming from octet 
($a_0$, $f_0$, $\kappa$) and singlet ($\sigma$) representations 
of $SU_V(3)$ group, constitute scalar nonet 
($\sigma$, $a_0$, $f_0$, $\kappa$). In order to study the chiral 
symmetry restoration at high temperatures, we will be investigating 
the trend of convergence in the masses of chiral partners occurring 
in pseudo scalar ($\pi$, $\eta$, $\eta'$, $K$) and scalar 
($\sigma$, $a_0$, $f_0$, $\kappa$) nonets, in the $2+1$ flavor 
symmetry breaking scenario.

\subsection{Polyakov Loop Potential }
\label{subsec:Plgtp}
The effective potential ${\cal U} \left( \Phi, \Phi^*, T \right)$ 
is constructed such that it reproduces thermodynamics of pure glue 
theory on the lattice for temperatures upto about twice the 
deconfinement phase transition temperature. In this work, we are 
using logarithmic form of Polyakov loop effective potential \cite{Ratti:07}. 
The results produced by this potential are known to be fitted well to the
lattice results. This potential is given by the following expression

\bea
\label{eq:logpot}
\frac{{\cal U_{\text{log}}}\left(\Phi,\Phi^*, T \right)}{T^4} &=& -\frac{a\left(T\right)}{2}\Phi^* \Phi +
b(T) \, \mbox{ln}[1-6\Phi^* \Phi \nn \\&&+4(\Phi^{*3}+ \Phi^3)-3(\Phi^* \Phi)^2]
\eea

where the temperature dependent coefficients are as follow

\begin{equation*} 
  a(T) =  a_0 + a_1 \left(\frac{T_0}{T}\right) + a_2 \left(\frac{T_0}{T}\right)^2 \; \; \; 
  b(T) = b_3 \left(\frac{T_0}{T}\right)^3\ .
\end{equation*}

The parameters 
of Eq.(\ref{eq:logpot}) are 
\begin{eqnarray*}
&& a_0 = 3.51\ , \qquad a_1= -2.47\ , \nn \\ 
&& a_2 = 15.2\ ,  \qquad  b_3=-1.75\ 
\end{eqnarray*}


The critical temperature for deconfinement phase transition 
$T_0=270$ MeV is fixed for pure gauge Yang Mills theory.
In the presence of dynamical quarks $T_0$ is directly linked to the
mass-scale $\Lambda$, the parameter which
has a flavor and chemical potential dependence in full dynamical QCD
and $T_0\to T_0(N_f,\mu)$. The $N_f$ and $\mu$ dependence of $T_0$ 
\cite{Schaefer:07,Braun,Herbst:2010rf,Pawl:Schaef,Schaef:12} is written as
\begin{equation}
  \label{eq:t0mu}
T_0(N_f, \mu) = T_{\tau} e^{-1/(\alpha_0 b(N_f, \mu))}
\end{equation}
where $T_\tau = 1.77$ GeV denotes the $\tau$ scale and
$\alpha_0=\alpha(\Lambda)$ the gauge coupling at some UV scale
$\Lambda$. The $\mu$-dependent running coupling reads
\begin{equation}
 b(N_f, \mu) = b(N_f) - b_{\mu}\frac{\mu^2}{ T_{\tau}^2}\ ,
\end{equation}
the factor $b_{\mu}\simeq \frac{16}{\pi}N_f$. Refs. \cite{Schaefer:07,Herbst:2010rf} contain 
the details of formula. Our present computations have been done
at $\mu=0$ and further since the $N_f$ dependence of $T_{0}$ has additional
complications of systematic error \cite{Herbst:2010rf}, we have taken $T_{0}$=270 MeV
in our calculation as in Ref. \cite{Schaef:12}.   
\section{Grand Potential in the Mean Field Approach}
\label{sec:potmf}
We are considering a spatially uniform system in thermal equilibrium at finite temperature
$T$ and quark chemical potential $\mu_f \ (f=u,\  d \ \text{and} \ s)$. The partition 
function is written as the path integral over quark/antiquark and meson 
fields \cite{gupta,Schaefer:09}
\bea
\label{eq:partf}
\mathcal{Z}&=& \mathrm{Tr\, exp}[-\beta (\hat{\mathcal{H}}-\sum_{f=u,d,s} 
\mu_f \hat{\mathcal{N}}_f)] \nn \\
&=& \int\prod_a \mathcal{D} \sigma_a \mathcal{D} \pi_a \int
\mathcal{D}q \mathcal{D} \bar{q} \; \mathrm{exp} \bigg[- \int_0^{\beta}d\tau\int_Vd^3x  \nn \\
&& \bigg(\mathcal{L_{QM}^{E}} 
 + \sum_{f=u,d,s} \mu_{f} \bar{q}_{f} \gamma^0 q_{f} \bigg) \bigg]. 
\eea
where $V$ is the three dimensional volume of the system,
$\beta= \frac{1}{T}$ and the superscript $\mathcal{E}$ denotes the euclidean 
Lagrangian. For three quark flavors, in general, the 
three quark chemical potentials are different. In this work, we 
assume that $SU_V(2)$ symmetry is preserved and neglect the small 
difference in masses of $u$ and $d$ quarks. Thus the quark chemical 
potential for the $u$ and $d$ quarks become equal $\mu_x = \mu_u = \mu_d$.
The strange quark chemical potential is $\mu_y = \mu_s$. Further we 
consider symmetric quark matter and net baryon number to be zero. 

	Here, the partition function is evaluated in the mean-field 
approximation \cite{Mocsy:01prc,Schaefer:08ax,gupta,Schaefer:09}. 
We replace meson fields by their expectation values 
$\langle M \rangle =  T_0 \bar{\sigma_0} + T_8 \bar{\sigma_8}$ 
and neglect both thermal as well as quantum fluctuations of meson 
fields while quarks and anti quarks are retained as quantum fields. 
Now following the standard procedure as given in 
Refs.~\cite{Kapusta_Gale,Schaefer:07,Ratti:06,Kfuku:04plb},
one can obtain the expression of grand potential as the sum of pure 
gauge field contribution ${\cal U} \left(\Phi, \Phi^*, T \right)$, 
meson contribution and quark/antiquark contribution evaluated in 
the presence of Polyakov loop,
\bea
\label{eq:grandp}
\Omega_{\rm MF}(T,\mu)=-\frac{T\ln Z}{V} &= &U(\sigma_x,\sigma_y)+{\cal U} \left(\Phi,\Phi^*,T \right) \nn \\
&&+ \Omega_{\bar{q}q} (T, \mu)
\eea 
The mesonic potential $U(\sigma_x,\sigma_y)$ is obtained from the $U(\sigma_0,\sigma_8)$ after 
transforming the original singlet-octet (0, 8) basis of condensates to
the non strange-strange basis $(x, y)$ as in Refs. \cite{Rischke:00,Schaefer:09,gupta,Schaef:12}.
We write the mesonic potential as
\bea
\label{eq:mesop}
&& U(\sigma_{x},\sigma_{y}) =\frac{m^{2}}{2}\left(\sigma_{x}^{2} +
  \sigma_{y}^{2}\right) -h_{x} \sigma_{x} -h_{y} \sigma_{y}
 - \frac{c}{2 \sqrt{2}} \sigma_{x}^2 \sigma_{y} \nn \\
 && + \frac{\lambda_{1}}{2} \sigma_{x}^{2} \sigma_{y}^{2}+
  \frac{1}{8}\left(2 \lambda_{1} +
    \lambda_{2}\right)\sigma_{x}^{4} +\frac{1}{8}\left(2 \lambda_{1} +
    2\lambda_{2}\right) \sigma_{y}^{4}\ 
\eea 
where 
\bea
\sigma_x &=&
\sqrt{\frac{2}{3}}\bar{\sigma}_0 +\frac{1}{\sqrt{3}}\bar{\sigma}_8, \\
\sigma_y &=&
\frac{1}{\sqrt{3}}\bar{\sigma}_0-\sqrt{\frac{2}{3}}\bar{\sigma}_8.
\eea
The chiral symmetry breaking external fields 
($h_x$, $h_y$) are written in terms of ($h_0$, $h_8$) analogously. 

Further the non strange and strange quark/antiquark decouple and the quark masses are 
\be 
m_x = g \frac{\sigma_x}{2}, \qquad m_y = g \frac{\sigma_y}{\sqrt{2}}
\ee 
Quarks become massive in symmetry broken phase because of non 
zero vacuum expectation values of the condensates.   
The  quark/antiquark contribution, in the presence of Polyakov loop potential, is written as 
\bea
&&\Omega_{\bar{q}q}(T,\mu)=\Omega_{q\bar{q}}^{\rm vac}+\Omega_{q\bar{q}}^{\rm T}=-2\sum_{f=u,d,s} \nn \\
&&\int\frac{d^3 p}{(2\pi)^3}\Big[{N_c E_f} \theta( \Lambda^2-\vec{p}^{\,2})+T \{ \ln g_{f}^{+} + \ln g_{f}^{-}\}\Big]
\label{Omeg_q}
\eea
The first term of the Eq.~(\ref{Omeg_q}) represents the fermion vacuum one loop
contribution, regularized by the ultraviolet cutoff $\Lambda$. The expressions $g_{f}^{+}$ and 
$g_{f}^{-}$ are defined  in the second term after taking trace over the color space
\bea
\label{eq:gpls} 
g_{f}^{+} = \Big[ 1 + 3\Phi e^{ -E_{f}^{+} /T} +3 \Phi^*e^{-2 E_{f}^{+}/T} +e^{-3 E_{f}^{+} /T}\Big] \,
\eea
\bea
\label{eq:gmns} 
g_{f}^{-} = \Big[ 1 + 3\Phi^* e^{ -E_{f}^{-} /T} +3 \Phi e^{-2 E_{f}^{-}/T} +e^{-3 E_{f}^{-} /T}\Big] \,
\eea
E$_{f}^{\pm} =E_f \mp \mu $ and $E_f$ is the
flavor dependent single particle energy of quark/antiquark and $m{_f}$ is the mass of the given quark flavor.
\be 
E_f = \sqrt{p^2 + m{_f}{^2}}
\ee 
\subsection{ The Renormalization Of Fermionic Vacuum Term  And The Effective Potential}
\label{subsec:Vterm}
The first term of Eq.~(\ref{Omeg_q}) can be properly renormalized  using the dimensional regularization scheme,
as done for two flavor case in Ref.\cite{Skokov:2010sf,Vivek:12,VKTR} and three flavor case in 
Ref.\cite{Schaef:12,Sandeep}. The brief description of essential steps are given in the following.
Fermion vacuum contribution is just the one-loop zero 
temperature effective potential at lowest order ~\cite{Quiros:1999jp}
\begin{eqnarray}
\label{vt_one_loop}
\Omega_{q\bar{q}}^{\rm vac} &=&-2 N_c \sum_{f=u,d,s} \int \frac{d^3
  p}{(2\pi)^3} E_f  \nonumber\\ 
&=&    - 2 N_c \sum_{f=u,d,s}  \int \frac{d^4 p}{(2\pi)^4}  \ln(p_0^2+E_{f}^{2})
+ {\rm K}
\end{eqnarray}
K is the infinite constant independent of
the fermion mass, hence it is dropped. The dimensional regularization of Eq.~(\ref{vt_one_loop}) near three 
dimensions, $d=3-2\epsilon$ gives the potential up to zeroth order in $\epsilon$ as
\begin{equation}
\Omega_{q\bar{q}}^{\rm vac} =\sum_{f=u,d,s}\frac{N_c \ m_f^4}{16 \pi^2} \left[
  \frac{1}{\epsilon} -\frac{ 
\{-3+2\gamma_E +4 \ln(\frac{m_f}{2\sqrt{\pi} \text{M}})\}}{2} \right]
\label{Omega_DR}
\end{equation}
Here M denotes the arbitrary renormalization scale.
The addition of a following counter term  $\delta \mathcal{L}$ to the QM/PQM model Lagrangian, 
\begin{equation}
\delta \mathcal{L} =\sum_{f=u,d,s} \frac{N_c}{16 \pi^2} m_f^4 \left[ \frac{1}{\epsilon} - \frac{1}{2}
 \left\{ -3 + 2 \gamma_E - 4 \ln (2\sqrt{\pi})\right\} \right] 
\label{counter}
\end{equation}
gives the renormalized fermion vacuum loop contribution as: 
\begin{equation}
\Omega_{q\bar{q}}^{\rm vac} =  -\sum_{f=u,d,s} \frac{N_c}{8 \pi^2} m_f^4  \ln\left(\frac{m_f}{\text{M}}\right)
\label{Omega_reg} 
\end{equation}
We note that the Polyakov loop potential and the temperature dependent part of the quark-antiquark contribution
to the grand potential in Eq.(\ref{eq:grandp}) vanishes at $T=0$ and $\mu=0$. The Polyakov loop order parameter $\Phi=\Phi^*$
becomes zero in the low temperature phase due to the phenomenon of color confinement and this makes the Polyakov loop 
potential ${\cal U_{\text{log}}}(\Phi,\Phi^*, T)$ zero at $T=0$ in Eq.(\ref{eq:logpot}).
The grand potential in vacuum becomes the renormalization scale M dependent when the fermionic 
vacuum loop contribution in the first term of Eq.(\ref{Omeg_q}), gets replaced by the appropriately renormalized term 
of Eq.(\ref{Omega_reg}) and we write:
\be
\Omega^{\rm M} (\sigma_x,\sigma_y) =U(\sigma_x,\sigma_y)+\Omega_{q\bar{q}}^{\rm vac}
\label{Omeg_rel}
\ee
The six unknown parameters $\text{m}^2$, $\lambda_1,\lambda_2$, $h_x,h_y$ and $c$ in the mesonic 
potential U($\sigma_x,\sigma_y$), are determined from the $\sigma_x$ and $\sigma_y$ dependent 
expressions of meson masses which are obtained by the double derivatives of the effective potential 
Eq.(\ref{Omeg_rel}) with respect to different meson fields. The mathematical details for determining
different parameters are given in the appendix A where the logarithmic M dependence of the term 
$\Omega_{q\bar{q}}^{\rm vac}$ gives rise to a renormalization scale M dependent part 
$\lambda_{2\text{M}}$ in the expression of the parameter $\lambda_2=\lambda_{2s}+n+\lambda_{2+}+\lambda_{\text{2M}}$. 
$\lambda_{2s}$ is the same old $\lambda_2$ parameter of the QM/PQM model in Ref.~\cite{Rischke:00,Schaefer:09,gupta}.
Here, $n=\frac{N_cg^4}{32\pi^2}$, $\lambda_{2+}=\frac{n{f_{\pi}}^2}{f_K \l f_K-f_{\pi}\r}\log\{\frac{2 f_K-f_{\pi}}{f_{\pi}}\}$
and $\lambda_{\text{2M}}= 4n\log\{ \frac{g\l 2f_K-f_{\pi}\r}{2 \text{M}}\}$.
After substituting this value of $\lambda_2$ in the expression of U($\sigma_x,\sigma_y$) 
and writing all the terms of summation in $\Omega_{q\bar{q}}^{\rm vac}$ expression explicitly, the Eq. (\ref{Omeg_rel}) can 
be rewritten as: 
\bea
\label{eq:mesVop}
&&\Omega^{\rm M} (\sigma_x,\sigma_y)=\frac{m^{2}}{2}\l\sigma_{x}^{2} +
\sigma_{y}^{2}\r -h_{x} \sigma_{x}-h_{y} \sigma_{y}
-\frac{c}{2 \sqrt{2}} \sigma_{x}^2 \sigma_{y} \nn \\
&& +\frac{\lambda_{1}}{4}\l\sigma_{x}^{4} +\sigma_{y}^{4}+2\sigma_{x}^{2}\sigma_{y}^{2}\r
+\frac{\l \lambda_{2\text{v}}+n+\lambda_{\text{2M}}\r}{8} \l\sigma_{x}^{4} + 2 \sigma_{y}^{4} \r \nn \\  
&&-\frac{n \sigma_{x}^4}{2}\log\l\frac{g\sigma_{x}}{2\text{M}}\r-\ n\sigma_{y}^4\log\l\frac{g \sigma_y}{\sqrt{2}\text{M}}\r   
\eea 
\begin{table*}[!ht]
 \begin{tabular}{|c||c|c|c|c|c|c|}
\hline
Model & c[MeV] & $m^2$ $[MeV^2]$  & $\lambda_1$ & $\lambda_{2s}$ & $h_x$ $[MeV^3]$& $h_y$ $[MeV^3]$ \\ \hline
QM W/$U_A(1)$& 4807.84 & $(342.52)^2$  &  1.40   & 46.48 & $(120.73)^3$ &$(336.41)^3$\\ \hline
QMVT W/$U_A(1)$& 4807.84 & $-(184.86)^2$ &-1.689 & 46.48 & $(120.73)^3$ &$(336.41)^3$\\ \hline
QM W/o$U_A(1)$&   0 &$-(189.85)^2$&  -17.01   & 82.47 & $(120.73)^3$ & $(336.41)^3$ \\ \hline 
QMVT W/o$U_A(1)$&  0  &$-(424.68)^2$&-20.46   & 82.47 & $(120.73)^3$ & $(336.41)^3$ \\ \hline
 \end{tabular}
\caption{parameters for $m_{\sigma} = 600$ MeV with and without $U_A(1)$
axial anomaly term.}
\label{tab:pmtr}
\end{table*}
here, $\lambda_{2\text{v}}=\lambda_{2s}+\lambda_{2+}$.
After rearrangement of terms, we find that
the logarithmic M dependence of $\lambda_2$ contained in $\lambda_{\text{2M}}$, completely cancels 
the scale dependence of all the terms in $\Omega_{q\bar{q}}^{\rm vac}$. The chiral part of the 
total effective potential now becomes free of any renormalization scale dependence. It is re expressed as
\bea
&& \Omega(\sigma_{x},\sigma_{y}) = \frac{m^{2}}{2}\l\sigma_{x}^{2} +
  \sigma_{y}^{2}\r -h_{x} \sigma_{x} -h_{y} \sigma_{y}
 - \frac{c}{2 \sqrt{2}} \sigma_{x}^2 \sigma_{y} \nn \\
 && + \frac{\lambda_{1}}{2} \sigma_{x}^{2} \sigma_{y}^{2}+
  \frac{\lambda_{1}}{4} \l \sigma_{x}^{4} +\sigma_{y}^{4} \r
    + \frac{ \l \lambda_{2\text{v}}+n\r}{8}\l \sigma_{x}^{4} + 2 \sigma_{y}^{4} \r \nn \\
 && -\frac{n \sigma_{x}^4}{2} \log\l \frac{\sigma_{x}}{\l2f_K-f_{\pi}\r}\r - n \sigma_{y}^4
    \log\l \frac{\sqrt 2\ \sigma_{y}}{\l2f_K-f_{\pi}\r}\r 
\label{eq:mesVop}
\eea
The calculation of vacuum meson masses from the effective potential also shows that the scale M dependence completely cancels out 
from their expressions. The explicit derivations of scale independent meson masses are given in the appendix B.

			In general $m_{\pi}$, $m_K$, the pion and kaon decay constant $f_\pi$, $f_K$,
mass squares of $\eta$, $\eta'$ and $m_\sigma$ are used to fix the six
parameters of the model.  The parameters are fitted such that in vacuum, 
the model produces observed pion mass $m_{\pi}$=138 MeV, kaon mass $m_K$= 496 MeV and $m_\eta'= 963(138)$ MeV,
 $m_\eta= 539(634.8)$ MeV for the case with the presence (absence, $c=0$) of axial anomaly 
term $c$. Numerical values of $\lambda_{2s}$ and $c$ are obtained easily after substituting the values of the
input parameters in their expressions in appendix A. Numerical values of $\lambda_{2+}$ and $n$ are obtained using
$f_\pi=92.4,f_K=113$ MeV and $N_{c}=3$. The scale independent expressions of $m^2_{\pi}$ and $m^2_{\sigma}$ 
given in the appendix B are exploited in the appendix A to obtain the vacuum values of the parameters
$m^2$ and $\lambda_1$  using $m_{\sigma}$=600 MeV. In the present work,
the $\lambda_{2s}$ and $c$ are the same as in the QM model \cite{Schaefer:09}, the value of $h_x$ and $h_y $
are also not affected by the fermionic vacuum correction. The parameters which are modified by the
fermionic vacuum correction are $m^2$, $\lambda_1$ and $\lambda_2$. Table \ref{tab:pmtr}
summarizes the numerical values of the parameters in different model scenarios. We point out that 
the effect of  one loop fermionic vacuum fluctuation in the 2+1 flavor renormalized
PQM model, has already been studied in the recent works of Refs. \cite{Schaef:12,Sandeep}. The model parameters 
($\lambda_2,\lambda_1$ and $m^2$) in these investigations are renormalization scale dependent and the cancellation
of scale dependence for the final results is achieved numerically.

  Now the thermodynamic grand potential in the presence of appropriately 
renormalized fermionic vacuum contribution in the Polyakov Quark Meson Model with 
vacuum term (PQMVT) model will be written as 
\bea
\Omega_{\rm MF}(T,\mu;\sigma_x,\sigma_y,\Phi,\Phi^*) & =& {\cal
U}(T;\Phi,\Phi^*) + \Omega(\sigma_x,\sigma_y) + \nn\\
&&\Omega_{q\bar{q}}^{\rm T}(T,\mu;\sigma_x, \sigma_y, \Phi,\Phi^*)
\label{OmegaMFPQMVT}
\eea
One can get the quark condensates $\sigma_x$, $\sigma_y$ and Polyakov
loop expectation values $\Phi$, $\Phi^*$ by searching the global 
minima of the grand potential for a given value of temperature $T$ 
and chemical potential $\mu$.
\be 
\label{eq:gapeq}
  \left.\frac{ \partial \Omega}{\partial
      \sigma_x} = \frac{ \partial \Omega}{\partial \sigma_y} 
      = \frac{ \partial \Omega}{\partial \Phi}
      = \frac{\partial \Omega}{\partial \Phi^*}
  \right|_{\sigma_x = \bar\sigma_x, \sigma_y=\bar\sigma_y, 
   \Phi=\bar\Phi, \Phi^* =\bar\Phi^*} = 0\ .
\ee 

\section{Meson Masses and Mixing Angles}
\label{sec:mixang}
	The curvature of the grand potential in Eq.(\ref{eq:grandp}) at the global 
minimum gives the finite temperature scalar and pseudo scalar meson masses.
\be 
\label{eq:sdergrand}
 {m_{\alpha,ab}^{2}} \bigg|_{T}=\frac{\partial^2 \Omega_{\rm MF} (T,\mu;\sigma_x, \sigma_y, \Phi,\Phi^*)}{\partial \xi_{\alpha,a} 
 \partial \xi_{\alpha,b}} \bigg|_{min}
\ee 
The subscript $\alpha=$ s, p ; s stands for scalar and p stands 
for pseudo scalar mesons and $a, b=0 \cdots 8$. 
\be 
\label{eq:MasFT}
{m_{\alpha,ab}^{2}} \bigg|_{T} =  m_{\alpha,ab}^{2} + ({\delta m^{\text{T}}_{\alpha,ab}})^{2}
\ee
The temperature dependence of meson masses comes from the temperature dependence of $\sigma_x$ and $\sigma_y$.
The term $({\delta m^{\text{T}}_{\alpha,ab}})^{2}$ results due to 
the explicit temperature dependence of quark-antiquark potential in  the grand potential. 
It vanishes in the vacuum where the meson mass matrix is 
determined as: 
\be
\label{eq:derVac}
 m_{\alpha,ab}^{2}=\frac{\partial^2 \Omega^{\rm M} (\sigma_x,\sigma_y)}{\partial \xi_{\alpha,a}
 \partial \xi_{\alpha,b}} \bigg|_{min}=({m^{\text{m}}_{\alpha,ab}})^{2}+({\delta m^{\text{v}}_{\alpha,ab}})^{2}
\ee
Here  the expressions $({m^{\text{m}}_{\alpha,ab}})^{2}$ as originally evaluated in Ref.~\cite{Rischke:00,Schaefer:09}, represent 
the second derivatives of the pure mesonic potential U($\sigma_x,\sigma_y$) at its minimum and the vacuum values 
of meson masses, $m_{\alpha,ab}^{2}$ , in the QM/PQM model are given only by these terms. The calculation details 
of mass modifications $({\delta m^{\text{v}}_{\alpha,ab}})^{2}$ resulting due to the fermionic vacuum correction, are presented
in the appendix A where we have also shown how those expressions are used for determining the model parameters. The Table IV 
of appendix A, contains all the expressions of $({m^{\text{m}}_{\alpha,ab}})^{2}$ and $({\delta m^{\text{v}}_{\alpha,ab}})^{2}$.  
The mass expressions $({m^{\text{m}}_{\alpha,ab}})^{2}$ have a renormalization scale M dependence in the QMVT/PQMVT model due
to the parameter $\lambda_{2}$. This dependence gets completely canceled by the already existing scale M dependence in the mass
modifications $({\delta m^{\text{v}}_{\alpha,ab}})^{2}$ and the final expressions of vacuum meson masses $m_{\alpha,ab}^{2}$,
are free of any renormalization scale dependence as shown explicitly in the appendix B.

In order to further calculate the in medium meson mass
modifications at finite temperature due to the quark-antiquark 
contribution in the presence of Polyakov loop potential, the complete 
dependences of all scalar and pseudo scalar meson fields in Eq.(\ref{eq:Mfld}) 
have to be taken into account. We have to diagonalize the resulting quark mass 
matrix. In the following, we recapitulate the expressions of mass modification due to the 
quark-antiquark contribution at finite temperature in the PQM model \cite{gupta} as:
\bea
\label{eq:ftmass}
&&({\delta m^{\text{T}}_{\alpha,ab}})^{2} \Big|_{PQM} = \frac{\partial^2\Omega_{q\bar{q}}^{\rm T} (T,\mu, \sigma_x, \sigma_y, \Phi, \Phi^*)}
 {\partial \xi_{\alpha,a} \partial \xi_{\alpha,b}} \Big|_{min} \nn \\
&&= 3 \sum_{f=x,y} \int \frac{d^3 p}{(2\pi)^3} \frac{1}{E_f}
 \biggl[ (A^{+}_{f} + A^{-}_{f} ) \biggl( m^{2}_{f,a b} - \frac{m^{2}_{f,a}
 m^{2}_{f, b}}{2 E_{f}^{2}} \biggl)  \nn \\
      && + (B_{f}^{+} + B_{f}^{-}) \biggl(  \frac{m^{2}_{f,a}  m^{2}_{f, b}}{2 E_{f} T}
\biggl) \biggl]  
\eea 
Here  $m^{2}_{f,a} \equiv \partial m^{2}_{f}/ \partial \xi_{\alpha,a}$ denotes
the first partial derivative and 
$m^{2}_{f,ab} \equiv \partial m^{2}_{f,a}/ \partial \xi_{\alpha,b}$ signifies
the second partial derivative of the squared quark mass with respect to the meson 
fields $\xi_{\alpha,b}$. These derivatives are evaluated in the Table III of Ref.
\cite{Schaefer:09}. We have given this table in the appendix A.
The notations $A_{f}^{\pm}$ and $B_{f}^{\pm}$ have the following 
definitions
\be
A_{f}^{+} =  \frac{\Phi e^{-E_{f}^{+}/T} + 2 \Phi^* e^{-2E_{f}^{+}/T} + e^{-3E_{f}^{+}/T}}{g_{f}^{+}}
\ee

\be
A_{f}^{-} =  \frac{\Phi^* e^{-E_{f}^{-}/T} + 2 \Phi e^{-2E_{f}^{-}/T} + e^{-3E_{f}^{-}/T}}{g_{f}^{-}}
\ee

and $B_{f}^{\pm}={3(A_{f}^{\pm}})^2 - C_{f}^{\pm}$, where we 
again define 

\be
C_{f}^{+} = \frac{\Phi e^{-E_{f}^{+}/T} + 4 \Phi^* e^{-2E_{f}^{+}/T} +3 e^{-3E_{f}^{+}/T}}{g_{f}^{+}}
\ee

\be
C_{f}^{-} = \frac{\Phi^* e^{-E_{f}^{-}/T} + 4 \Phi e^{-2E_{f}^{-}/T} +3 e^{-3E_{f}^{-}/T}}{g_{f}^{-}}
\ee

In the PQMVT model, the final expression  for finite temperature meson masses in Eq. (\ref{eq:MasFT}) is written as
\be 
{m_{\alpha,ab}^{2}} \bigg|_{T,PQMVT} =  m_{\alpha,ab}^{2} + ({\delta m^{\text{T}}_{\alpha,ab}})^{2} \bigg|_{PQM}
\ee
This expression gives meson masses in the PQM model also when the fermionic vacuum contribution becomes zero in the expression of 
vacuum meson masses in the first term. The expression for the finite temperature meson mass modifications $({\delta m^{\text{T}}_{\alpha,ab}})^{2}
|_{QM}$ due to the quark-antiquark potential in the QM model, can be found in Ref. \cite{Schaefer:09}. We use this expression to
write the finite temperature meson masses in the QMVT model as
\be 
{m_{\alpha,ab}^{2}} \bigg|_{T,QMVT} =  m_{\alpha,ab}^{2} + ({\delta m^{\text{T}}_{\alpha,ab}})^{2} \bigg|_{QM}
\ee
Here also, the same expression gives meson masses in the QM model when the fermionic vacuum correction is absent in the expression of vacuum meson masses.
 
	The diagonalization of (0,8) component 
of mass matrix gives the masses of $\sigma$ and $f_0$ mesons in scalar 
sector and the masses of $\eta'$ and $\eta$ in pseudo scalar sector. 
The scalar mixing angle $\theta_s$ and pseudo scalar mixing angle 
$\theta_p$ are given by
\be 
\tan{2\theta_{\alpha}} = \biggl( \frac{2 m^2_{\alpha,08}}
{m^2_{\alpha,00}-m^2_{\alpha,88}} \biggr)
\ee
The appendix C of Ref.\cite {Schaefer:09} contains all the transformation details of the mixing for the (0,8) basis
that generates the physical basis of the scalar ($\sigma$,$f_0$) and pseudo-scalar ($\eta'$, $\eta$) mesons. This appendix
also explains the ideal mixing, and gives the details of formulae by which the physical mesons transform into the mesons which are 
pure strange or non-strange quark systems. 
\section{Fermionic Vacuum Correction and Chiral  Restoration}
\label{sec:eploop}

We are investigating the effect of fermionic vacuum fluctuation on the restoration 
of chiral symmetry when it is properly accounted for in the $2+1$ flavor quark meson
model and PQM model at finite temperature and zero chemical potential with and without
 axial $U_A(1)$ breaking. We have compared the results of present computations in the QMVT and 
PQMVT models with the already existing calculations in the quark meson model and PQM
model \cite{gupta,Schaefer:09}. The interplay of the effect of $U_A(1)$ axial 
restoration and chiral symmetry restoration in the influence of fermionic vacuum 
fluctuation has been investigated and compared with in different model scenarios through 
the temperature variation of strange, non strange chiral condensates, meson masses 
and mixing angles. The $U_A(1)$ axial breaking term is constant throughout the computation. 
The value of Yukawa coupling $g= 6.5$ has been fixed from the non strange constituent quark mass 
$m_q = 300$ MeV in vacuum ($T=0, \mu=0)$. This predicts the vacuum strange quark mass 
$m_s \backsimeq 433$ MeV

\begin{figure*}[!htbp]
\subfigure[Non-strange condensate variation with $U_A(1)$ anomaly term]{
\label{fig:mini:fig1:a} 
\begin{minipage}[b]{0.45\linewidth}
\centering \includegraphics[width=3.4in]{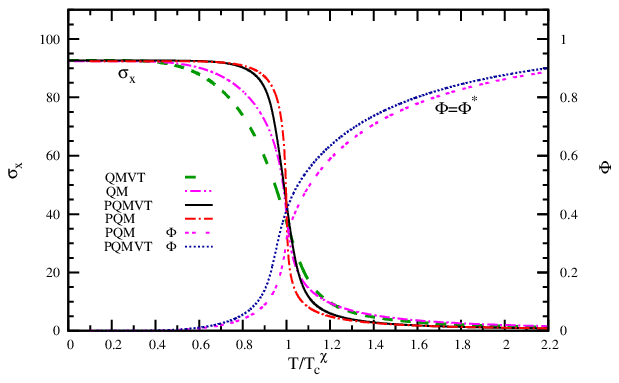}
\end{minipage}}%
\hspace{.15in}
\subfigure[Strange condensate variation with and without $U_A(1)$ anomaly term]{
\label{fig:mini:fig1:b} 
\begin{minipage}[b]{0.45\linewidth}
\centering \includegraphics[width=3.4in]{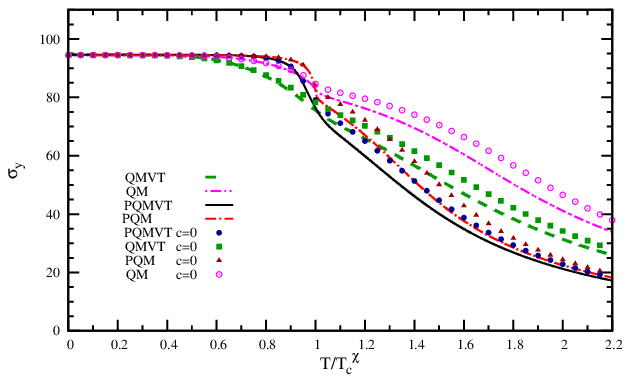}
\end{minipage}}
\caption{The Fig.\ref{fig:mini:fig1:a}, shows the reduced temperature scale ($T/T_{c}^{\chi}$) 
variation of the non strange condensate $\sigma_x$ at zero chemical potential($\mu = 0$) and non
zero axial anomaly ($c \neq 0$) in the QM,QMVT,PQM and PQMVT model calculations. 
The dash double dots line in magenta, the thick long dash line in dark green and the dash dot line in red ,
represent the respective $\sigma_x$ variations in the QM, QMVT and PQM model while the
solid black line represents the PQMVT model $\sigma_x$ variation. The same line types in the
Fig.\ref{fig:mini:fig1:b}, represent the respective model variations of strange  condensate $\sigma_y$
when $c \neq 0$. The line of solid circle dots in dark blue  and the line of solid triangle
dots in deep red  in the Fig.\ref{fig:mini:fig1:b}, show the respective variations of the $\sigma_y$ 
in the PQMVT and PQM models while the line of solid square dots in green and the line of hollow circle
dots in magenta represent the respective  $\sigma_y$ variations in the QMVT and QM models when axial 
anomaly term is absent i.e. $c=0$. The expectation value of the Polyakov loop field $\Phi$, is shown in the 
right side plots of Fig.\ref{fig:mini:fig1:a} where dot like small dash line in dark blue represents
the $\Phi$ variation in the PQMVT model while the double dash line in magenta represents the 
$\Phi$ variation in the PQM model.}
\label{fig:mini:fig1} 
\end{figure*}
\subsection{Condensates And Fermionic Vacuum Correction }
\begin{table}[!htb]
\begin{tabular}{|c|c|c|c|c|}
\hline
& QM & QMVT & PQM & PQMVT  \\
\hline
$T_{c}^{\chi}$(MeV)& $146.1$ & $171.1$ & $205.8$ & $216.5$ \\
$T_{s}^{\chi}$(MeV)&$248.3\pm2.0$&$247.8\pm2.5$&$274\pm1.5$&$269.\pm4.0$\\
$T_{c}^{\Phi}$ (MeV) & $ -  $ & $ - $ & $205.6$ &$205.6$ \\
\hline
\end{tabular}
\caption{The table of characteristic temperature (pseudo critical temperature) 
for the chiral transition in the non-strange sector $T_{c}^{\chi}$, 
strange sector $T_{s}^{\chi}$ and the confinement-deconfinement transition $T_{c}^{\Phi}$,
in the QM, QMVT, PQM and PQMVT model. $\pm$ gives the temperature range near  $T_{s}^{\chi}$  over which the
rather flat and broad second peak of the strange condensate derivative $\frac{\partial\sigma_y}{\partial T}$, 
shows a distinct change of about 0.1 percent of the numerical value of the second peak height. }
\label{tab:ctemp}
\end{table}
	The solutions of the gap equations Eq.(\ref{eq:gapeq})  at zero 
chemical potential, yield the temperature dependence of the Polyakov loop expectation value
$\Phi$, non strange and strange condensates and the inflection point of these order 
parameters respectively give the characteristic temperature 
(pseudo-critical temperature) for the confinement - deconfinement 
transition $T_{c}^{\Phi}$, the chiral transition in the non-strange 
$T_{c}^{\chi}$ and strange sector $T_{s}^{\chi}$. Table \ref{tab:ctemp} shows
the various pseudo-critical temperatures in different models. We will use a reduced 
temperature scale $T/T_{c}^{\chi}$ to compare the PQMVT(QMVT) model variations with 
that of the PQM(QM) model because the absolute comparison of the characteristic temperatures 
between two models of the same universality class can not be 
made according to the Ginsburg-Landau effective theory \cite{Costa:09}.

For $T=0$, the Fig.\ref{fig:mini:fig1:a} shows that the condensate $\sigma_x$ = 92.4 MeV while 
the $\sigma_y$ =94.5 MeV in Fig.\ref{fig:mini:fig1:b}. The $U_A(1)$ anomaly ($c \neq 0$) has a negligible 
effect on the non-strange condensate $\sigma_x$ variation which is sharpest for the $T/T_{c}^{\chi}$ = 0.9 
to 1.2 range in the PQM model. The $\sigma_x$ variation  becomes 
smoother in the PQMVT model on account of the fermionic vacuum correction and its most smooth
variation results in the  QMVT model due to the absence of Polyakov loop potential.
The fermionic vacuum correction together with the Polyakov loop potential gives rise to a
largest degree of strange condensate $\sigma_y$  melting in the PQMVT model when $c \neq 0$ 
in Fig.\ref{fig:mini:fig1:b}. The PQM model $\sigma_y$  melting is already reported \cite{gupta}
to be significantly larger than that of the QM model. The effect of only the fermionic vacuum correction
is quite robust as evident from a noticeably larger  melting of the $\sigma_y$ in the QMVT model. Comparing the
model results of the $\sigma_y$ temperature variation for the $c=0$  case with that of the  $c \neq 0$ case in Fig.\ref{fig:mini:fig1:b},
we conclude that the melting of the strange condensate gets  reduced in the same small proportion in all the models when the 
axial anomaly term is absent.
\begin{figure*}[!htbp]
\subfigure[ With $U_A(1)$ anomaly term]{
\label{fig:mini:fig2:a} 
\begin{minipage}[b]{0.45\linewidth}
\centering \includegraphics[width=3.40in]{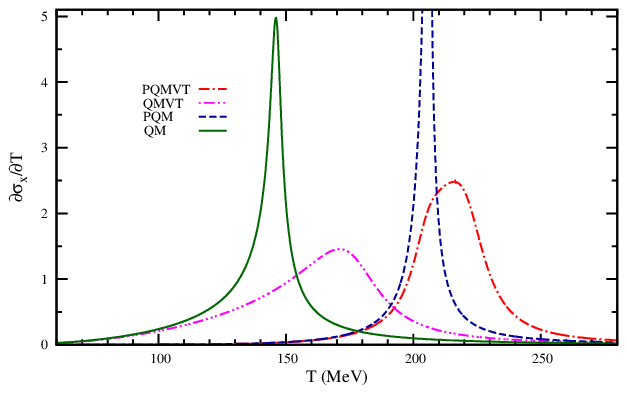}
\end{minipage}}%
\hspace{0.10in}
\subfigure[ With $U_A(1)$ anomaly term]{
\label{fig:mini:fig2:b} 
\begin{minipage}[b]{0.45\linewidth}
\centering \includegraphics[width=3.40in]{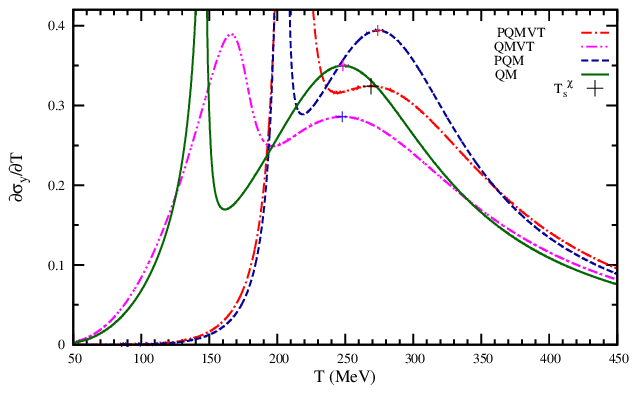}
\end{minipage}}
\caption{Fig.\ref{fig:mini:fig2:a} shows the temperature variation of the
$\frac{\partial \sigma_x}{\partial T}$. The dash dot in red, dash double dot in magenta 
and solid line in dark green  show the respective PQMVT, QMVT and QM model variations 
with their distinct peaks. The dash line in dark blue shows the PQM model variation
whose very high peak is not visible on the y-axis scale which has been chosen such as to highlight
the peaks for other model variations in the Fig.\ref{fig:mini:fig2:a}. The same line types
in the Fig.\ref{fig:mini:fig2:b}, show the respective model temperature variations of the strange
condensate temperature derivative $\frac{\partial \sigma_y}{\partial T}$. This variation shows two
peaks where the first peak is caused by the chiral dynamics in the non-strange sector. The location of the second peak
( marked by the plus symbol in the Fig.\ref{fig:mini:fig2:b}) gives strange sector 
chiral crossover transition temperature $T_{s}^{\chi}$. The second peak is very broad and flat over a small temperature
range and its location is marked by an ambiguity range of $\pm$ for the $T_{s}^{\chi}$ (given in the Table \ref{tab:ctemp})
in which the derivative $\frac{\partial\sigma_y}{\partial T}$, shows a distinct change of about 0.1 percent
of the numerical value of the second peak height.}
\label{fig:mini:fig2} 
\end{figure*}

	Curves ending in the right side of the Fig.\ref{fig:mini:fig1:a}, represent the 
temperature variation of the Polyakov loop expectation value 
$\Phi$. Since $\mu=0$ in our calculations, we have $\Phi$=$\Phi^*$.  
Here we recall that the improved ansatz of the logarithmic polyakov loop
potential \cite{Kaczmarek:02,Ratti:07,Hansen:07,Kfuku:04plb} avoids 
the $\Phi$ expectation value higher than one and hence describes the dynamics of gluons more 
effectively.

The peak in the temperature variation of $\frac{\partial\sigma_x}{\partial T}$ in Fig.\ref{fig:mini:fig2:a} 
gives the $T_{c}^{\chi}$ for the chiral crossover at $\mu=0$.
It is evident from the plots in Fig.\ref{fig:mini:fig2:a} and the values given in Table \ref{tab:ctemp}
that the fermionic vacuum correction causes a smoother and gentler crossover transition in the non-strange sector
where the transition temperature $T_{c}^{\chi}$ for the  PQMVT(QMVT) model increases by 10.7(25) MeV over its 
PQM(QM) model value. The confinement-deconfinement crossover transition temperature $T_{c}^{\Phi}$=205.6 MeV is
same in both the models PQM and PQMVT. But unlike the PQM model, the deconfinement transition for the PQMVT model,
does not remain coincident with the non-strange sector chiral crossover transition and we get $T_{c}^{\chi}>T_{c}^{\Phi}$.
The chiral crossover is coincident with the confinement-deconfinement transition in the RBC-Bielefeld and HotQCD lattice 
calculations where $T_{c}^{\chi}$ lies between 185-195 MeV \cite{HotQCD,ChengLQCD,BazLQCD} but the Wuppertal-Budapest(WB)
Collaboration in comparison gives a pseudo-critical temperature which is 40 MeV smaller for the non-strange crossover
transition  and 15 MeV smaller for the deconfinement transition and $T_{c}^{\chi}<T_{c}^{\Phi}$ \cite{WBLQCD,LQCDWB2,LQCDWBL}.
In Our PQMVT model calculation, we have taken $m_{\sigma}$=600 and $T_0$=270 MeV in order to compare results with the 
earlier work done in the QM and PQM models in Ref. \cite{Schaefer:09,Schaefer:08ax,gupta}.
This choice does not reproduce the  Wuppertal-Budapest scenario and is more in tune with the standard scenario of the PQM
model calculations of Schaefer et. al. in Ref.\cite{Schaefer:09ax} where they have done detail comparisons of various transitions
with different parameter sets and three different parametrization of the Polyakov loop potential. The recent
HotQCD lattice results show smaller disagreement in  the transition temperature value \cite{HotLQCDL} when compared with 
the WB results for the physical pion mass. We point out that most lattice calculations are carried out with periodic boundary condition, which
is convenient for the computations, but rather far from the experimental setup. An exploratory quenched study \cite{quench} suggests 
that critical temperatures with realistic boundary conditions can be up to 30 MeV larger than the values, which are 
measured in conventional lattice calculations. In effective model investigations, the $T_{c}^{\chi}$ and $T_{c}^{\Phi}$  
values are quite sensitive to the chosen models and  parameter sets. In the NJL and PNJL model investigations \cite{Hiller}, for example, 
the consideration of the eight quark interactions, leads to the significant  lowering of the pseudo-critical temperature for the
chiral crossover transition. Further the smaller values of $T_{0}$ for the Polyakov loop potential also leads to considerable lowering of the transition temperature \cite{Ratti:06}. The parameters corresponding to $m_{\sigma}$=400 MeV in our PQMVT model calculation 
give,  $T_{c}^{\chi}$=202.6 MeV and $T_{c}^{\Phi}$=201.1 MeV. Here the non-strange chiral crossover and deconfinement transitions
are almost coincident similar to the recent results of Schaefer et. al. \cite{Schaef:12} in the renormalized PQM model.
\begin{figure*}[!htbp]
\subfigure[PQMVT and PQM model results:axial anomaly term $c\neq0$]{
\label{fig:mini:fig3:a} 
\begin{minipage}[b]{0.45\linewidth}
\centering \includegraphics[width=3.4in]{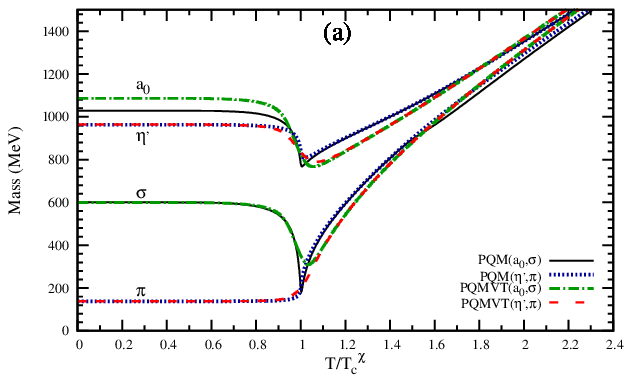}
\end{minipage}}%
\hspace{0.10in}
\subfigure[QMVT and QM model results:axial anomaly term $c\neq0$]{
\label{fig:mini:fig3:b} 
\begin{minipage}[b]{0.45\linewidth}
\centering \includegraphics[width=3.4in]{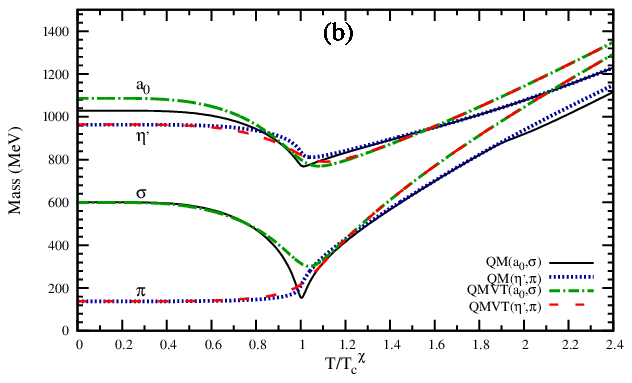}
\end{minipage}}
\caption{Mass variations for the chiral partners ($\sigma$, $\pi$) and ($a_0$, $\eta'$) 
on the reduced temperature ($T/T_{c}^{\chi}$) scale at $\mu=0$,
are plotted in Fig.\ref{fig:mini:fig3:a} for the PQMVT and PQM model and the corresponding mass
variations in the QMVT and QM model, are plotted in Fig.\ref{fig:mini:fig3:b}. The  dash dot line
plots in dark green  and the solid line plots in black respectively for the PQMVT(QMVT) 
model and the PQM(QM) model show the $\sigma$ and $a_0$ mass variations in the left panel(right panel).
The $\pi$ and $\eta'$ mass variations are denoted by the dash line in red  plots and the thick dots 
line in blue plots respectively for the  PQMVT(QMVT) model and the QMVT(QM) model in the 
left panel(right panel).}
\label{fig:mini:fig3} 
\end{figure*}
\begin{figure*}[!htbp]
\subfigure[PQMVT and PQM model results:axial anomaly term $c=0$]{
\label{fig:mini:fig4:a} 
\begin{minipage}[b]{0.45\linewidth}
\centering \includegraphics[width=3.4in]{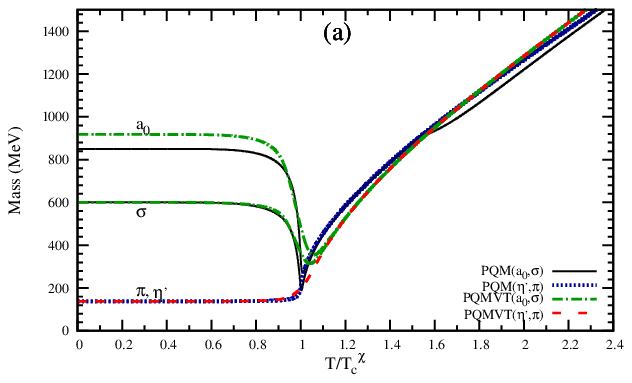}
\end{minipage}}%
\hspace{0.10in}
\subfigure[QMVT and QM model results:axial anomaly term $c=0$]{
\label{fig:mini:fig4:b} 
\begin{minipage}[b]{0.45\linewidth}
\centering \includegraphics[width=3.40in]{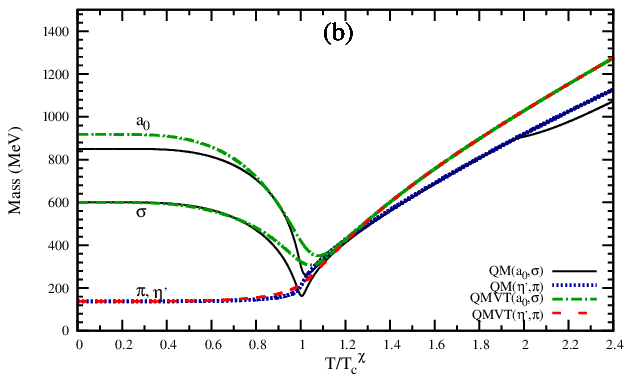}
\end{minipage}}
\caption{The line types in Fig.\ref{fig:mini:fig4:a} and 
in Fig.\ref{fig:mini:fig4:b} represent the same mass variations as 
depicted in Fig.\ref{fig:mini:fig3} but here in these computations
the axial $U_A(1)$ anomaly term is absent i.e. $c=0$.}
\label{fig:mini:fig4} 
\end{figure*}
															
The chiral crossover transition in the strange sector is lot more smooth and weaker than
the crossover transition of non-strange sector for all the models due to the large 
constituent mass of the strange quark $m_s$ = 433 MeV in vacuum. 
The variation of the temperature derivative of
$\sigma_y$ shows two peaks in Fig.\ref{fig:mini:fig2:b} for all the models, the first peak is higher and sharper because it 
is driven by the chiral crossover transition dynamics in the non-strange sector. The crossover 
temperature $T_{s}^{\chi}$ in the strange sector is identified in Fig.\ref{fig:mini:fig2:b} by locating the position of
the second peak which is quite broad, smooth and flat over a small temperature range in all 
the models. The ambiguity in the identification of second peak (marked by the plus symbol in the Fig.\ref{fig:mini:fig2:b})
is indicated by the $\pm$ flatness range for $T_{s}^{\chi}$ in the Table \ref{tab:ctemp}.  
The largest but smoother melting of strange condensate is obtained in the PQMVT model 
with $T_{s}^{\chi}$= $269.0\pm4$ MeV. It will have an interesting physical consequence 
in the early setting up of a smoother mass degeneration trend in  masses of the
chiral partners ($K$, $\kappa$) and ($\eta$, $f_0$) and in the early emergence of a smoother $U_A(1)$ restoration trend. 
\subsection{ Meson Mass Variations }

\begin{figure*}[!htbp]
\subfigure[PQMVT and PQM model results:axial anomaly term $c\neq0$]{
\label{fig:mini:fig5:a} 
\begin{minipage}[b]{0.45\linewidth}
\centering \includegraphics[width=3.40in]{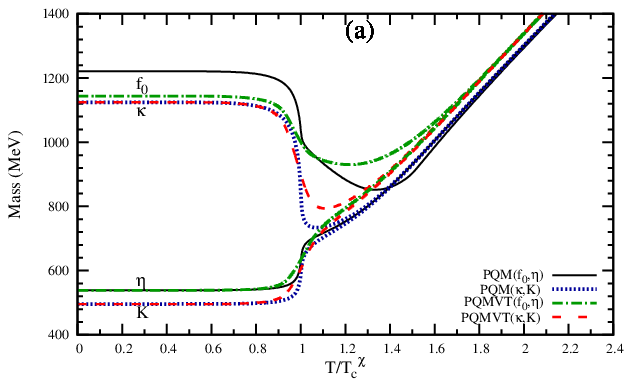}
\end{minipage}}%
\hspace{0.10in}
\subfigure[QMVT and QM model results:axial anomaly term $c\neq0$]{
\label{fig:mini:fig5:b} 
\begin{minipage}[b]{0.45\linewidth}
\centering \includegraphics[width=3.40in]{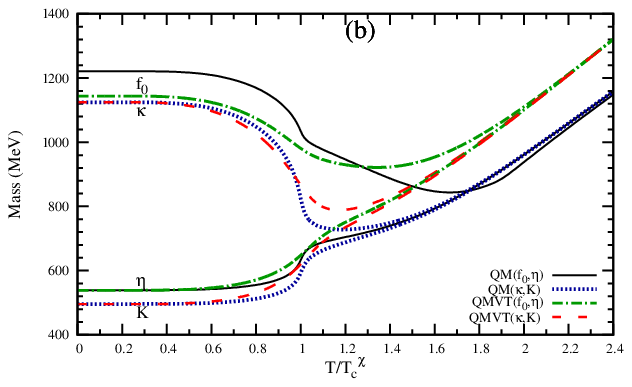}
\end{minipage}}
\caption{Mass variations for the chiral partners ($\eta$, $f_0$) and ($K$, $\kappa$) 
on the reduced temperature ($T/T_{c}^{\chi}$) scale at $\mu=0$,
are plotted in Fig.\ref{fig:mini:fig5:a} for the PQMVT and PQM model and the corresponding mass
variations in the QMVT and QM model, are plotted in Fig.\ref{fig:mini:fig5:b}. The  dash dot line
plots in dark green  and the solid line plots in black respectively for the PQMVT(QMVT) 
model and the PQM(QM) model show the $\eta$ and $f_0$ mass variations in the left panel(right panel).
The $K$ and $\kappa$ mass variations are denoted by the dash line red plots and the thick dots 
line in blue plots respectively for the  PQMVT(QMVT) model and the PQM(QM) model in the 
left panel(right panel).}
\label{fig:mini:fig5} 
\end{figure*}

\begin{figure*}[!htbp]
\subfigure[PQMVT and PQM model results:axial anomaly term $c=0$]{
\label{fig:mini:fig6:a} 
\begin{minipage}[b]{0.45\linewidth}
\centering \includegraphics[width=3.40in]{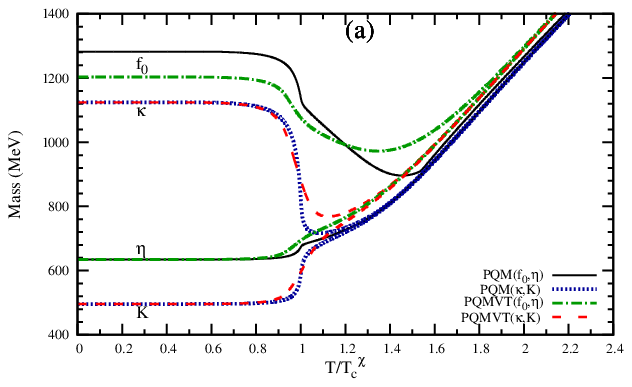}
\end{minipage}}%
\hspace{0.10in}
\subfigure[QMVT and QM model results:axial anomaly term $c=0$]{
\label{fig:mini:fig6:b} 
\begin{minipage}[b]{0.45\linewidth}
\centering \includegraphics[width=3.40in]{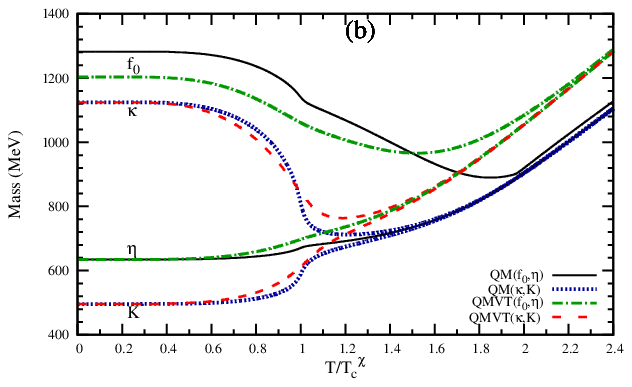}
\end{minipage}}
\caption{ The line types in Fig.\ref{fig:mini:fig6:a} and 
in Fig.\ref{fig:mini:fig6:b} represent the same mass variations as 
depicted in Fig.\ref{fig:mini:fig5} but here in these computations
the axial $U_A(1)$ anomaly term is absent i.e. $c=0$.}
\label{fig:mini:fig6} 
\end{figure*}

The meson mass temperature variations of the PQMVT(QMVT) model in the presence of axial 
$U_A(1)$ anomaly term, are compared with the corresponding PQM(QM) model results
respectively in the Fig.\ref{fig:mini:fig3:a} (Fig.\ref{fig:mini:fig3:b})
for the chiral partners ($\sigma$,$\pi$) and ($a_0$,$\eta'$) and the Fig.\ref{fig:mini:fig5:a} (Fig.\ref{fig:mini:fig5:b})
for the chiral partners ($\eta$,$f_0$) and ($K$,$\kappa$). The analogous plots of mass variations when the  axial 
$U_A(1)$ anomaly term ($c=0$) is absent, are given respectively in the Fig.\ref{fig:mini:fig4:a} (Fig.\ref{fig:mini:fig4:b}) 
and Fig.\ref{fig:mini:fig6:a} (Fig.\ref{fig:mini:fig6:b}).

The sharpest mass degeneration of the PQM model for the ($\sigma$,$\pi$) and ($a_0$,$\eta'$) mesons, becomes quite 
smooth in the Fig.\ref{fig:mini:fig3:a} for the PQMVT model due to the smoother melting of the non-strange condensate
$\sigma_x$ caused by the fermionic vacuum correction in the Fig.\ref{fig:mini:fig1:a}. The most smooth mass degeneration 
results are in Fig.\ref{fig:mini:fig3:b} for the QMVT model variations because the Polyakov loop effect
which causes a sharper chiral crossover transition is absent. Similar trend of smoother mass degeneration is seen in 
the masses of the chiral partners ($\eta$,$f_0$) and ($K$,$\kappa$) in the Fig.\ref{fig:mini:fig5:a}(Fig.\ref{fig:mini:fig5:b})
for the PQMVT(QMVT) model. Since a significant melting of the strange condensate  $\sigma_y$ occurs at a higher 
temperature in all the models, the $K$, $\kappa$ and $\eta$ meson masses become degenerate not exactly at 
$T/T_c^{\chi}=1$ but around $T/T_c^{\chi}=1.3 (1.2)$ in the PQMVT(PQM) model.
The $f_0$ meson mass intersects the degenerate line of $K$, $\kappa$ and $\eta$  meson masses  around 
$T/T_c^{\chi}=1.4(1.8)$ in the Fig.\ref{fig:mini:fig5:a}(Fig.\ref{fig:mini:fig5:b}) for the PQM(QM) model computations, and 
then becomes smaller than the $m_{\eta}$ developing a kink like structure after crossing it. Later, the $f_0$ meson mass 
degenerates with the $m_K$, $m_\kappa$ and $m_\eta$ variations again for $T/T_c^{\chi}>1.8(2.3)$. We find that the kink
in the $f_0$ meson mass variation altogether disappears from the PQMVT(QMVT) model results
in the Fig.\ref{fig:mini:fig5:a}(Fig.\ref{fig:mini:fig5:b}) due to the robust 
effect of the fermionic vacuum correction and the  $m_{f_0}$  degenerates quite smoothly with the $m_K$, $m_\kappa$ and $m_\eta$
earlier at $T/T_c^{\chi}>1.7(1.9)$ and remains so forever. These  mass degeneration trends
reflect the effect of fermionic vacuum fluctuation on the chiral symmetry restoration in the strange sector and  
result due to the smoother but larger (largest in the PQMVT model) melting of the strange condensate in Fig.\ref{fig:mini:fig1:b}.

	The PQM (QM) model $\sigma$ meson mass variation also shows a kink structure  which starts at 
$T/T_c^{\chi}=1.5\ (1.9)$ and persists afterwards in the Fig.\ref{fig:mini:fig3:a} (Fig.\ref{fig:mini:fig3:b}). 
This kink again disappears from the  $m_\sigma$ variations in the PQMVT and QMVT model where
the smooth line of the degenerated $m_\sigma$ and $m_\pi$ in the Fig.\ref{fig:mini:fig3:a} and  Fig.\ref{fig:mini:fig3:b}, 
show closer convergence towards the degenerate masses of $a_0$ and $\eta'$ mesons for higher $T/T_c^{\chi}>1$ and the mass 
gap between these two sets of chiral partners, becomes small in comparison to the mass gap seen in the PQM and QM model.
Here we recall that the  $U_A(1)$ axial symmetry breaking, generates the mass gap between the two sets of 
the chiral partners, ($\sigma$, $\pi$) and ($a_0$, $\eta'$) i.e. $m_{\pi} = m_{\sigma}$ $<$ $m_{a_0} = m_{\eta'}$ for $T/T_C^{\chi}>1$
because the anomaly term ($\frac{c \ \sigma_y}{\sqrt{2}}$) has opposite sign in the expressions of 
$m_{a_0}$ and  $m_{\pi}$ \cite{Schaefer:09}. Hence the mass gap reduction will be larger due to the larger melting of $\sigma_y$ for higher $T/T_c^{\chi}$ in the PQMVT model. 
We thus conclude that apart from effecting the smoother occurrence of chiral 
$SU_L(2) \times SU_R(2)$ symmetry restoration in the non-strange sector,
the inclusion of fermionic vacuum fluctuation in the PQM (QM) model also effects an 
early and smoother set up of the $U_A(1)$ restoration trend.

\begin{figure*}[!htbp]
\subfigure[Scalar and Pseudo scalar mixing angle variations in PQMVT and PQM model]{
\label{fig:mini:fig7:a} 
\begin{minipage}[b]{0.45\linewidth}
\centering \includegraphics[width=3.40in]{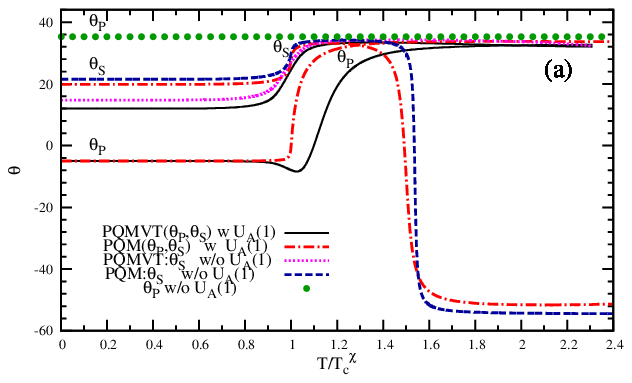}
\end{minipage}}%
\hspace{0.10in}
\subfigure[Scalar and Pseudo scalar mixing angle variations in PQM and QM model]{
\label{fig:mini:fig7:b} 
\begin{minipage}[b]{0.45\linewidth}
\centering \includegraphics[width=3.40in]{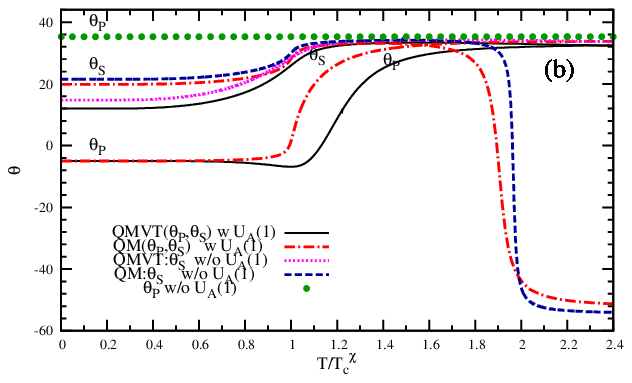}
\end{minipage}}
\caption{In the presence of axial anomaly, the lower(upper) black solid line 
and dash dot line in red  depict the $\theta_P$($\theta_S$)
variations respectively in the PQMVT and PQM model computations in 
the Fig.\ref{fig:mini:fig7:a} while the same line types represent the corresponding
variations respectively for the QMVT and QM model in the Fig.\ref{fig:mini:fig7:b}.
In the absence of axial anomaly i.e. $c=0$, the dot like small dash line in magenta and 
dash line in dark blue represent the scalar $\theta_S$ mixing angle variations 
respectively in the PQMVT and PQM model computations in the Fig.\ref{fig:mini:fig7:a}
while the same line types represent the corresponding variations respectively for the 
QMVT and QM model in the Fig.\ref{fig:mini:fig7:b}. The pseudo scalar 
$\theta_P$ mixing angle variations for $c=0$ are constant and are shown by the dark green
filled circular dots for both the PQMVT and PQM model 
calculations in the Fig.\ref{fig:mini:fig7:a} while the same line type represents the $\theta_P$
variations for the QMVT and QM model in the Fig.\ref{fig:mini:fig7:b}.} 
\label{fig:mini:fig7} 
\end{figure*}
In the absence of the axial anomaly $c=0$, the $m_{\eta'}$ always remains equal to $m_{\pi}$ and the
mass degeneration of the chiral partners ($\sigma$, $\pi$) and ($a_0$, $\eta'$) results near $T/T_c^{\chi}=1.0$ 
in all the model plots. Here also the PQM (QM) model prominent kink structure, which forms near $T/T_c^{\chi}=1.5(1.9)$ in
the $m_\sigma$ variation in Fig.\ref{fig:mini:fig4:a}(Fig.\ref{fig:mini:fig4:b}), gets completely smoothed out in the PQMVT(QMVT) 
model. Further, in Fig.\ref{fig:mini:fig6:a} (Fig.\ref{fig:mini:fig6:b}), 
the $m_{f_0}$ variation in the PQM (QM) model for the $c=0$ case, does not become completely degenerate with 
the $m_{\eta}$ though it becomes very close (nearly touches) to the $\eta$ mass variation when $T/T_c^{\chi} \sim 1.6(2.0)$ 
and afterwards $m_{f_0}$ takes slightly larger value than the $m_{\eta}$.
The $f_0$ mass variation, in contrast, degenerates quite smoothly with the $m_{\eta}$ when $T/T_c^{\chi} \sim 1.9(2.3)$ in the 
PQMVT (QMVT) model. Thus the fermionic vacuum correction leads to the smoother mass degeneration trends also when $c=0$. 

 Here we mention another noteworthy result. In the influence of the fermionic vacuum correction, the scalar particle vacuum mass
increases to 1086.26(917.93) MeV for the $a_{0}$ meson and decreases to 1143.92(1203.16) MeV for the $f_{0}$ meson in
the presence(absence) of axial anomaly in the QMVT/PQMVT model from the respective vacuum mass value of $m_{a_{0}}$=1028.7(850.5) MeV
and $m_{f_{0}}$=1221.1(1282.3) MeV in the QM/PQM model. Further, we point out that the kinks in the PQM/QM model $m_\sigma$ and $m_{f_0}$  
variations, are the consequence of an interchange in their identities for higher values on the reduced temperature 
scale \cite{gupta, Schaefer:09}. Here, we again emphasize that the crossing 
or anti-crossing pattern in the meson mass variations, completely disappears when the fermionic vacuum 
fluctuation is accounted for in the PQMVT and QMVT model. In order to have a 
proper perspective of the  PQM/QM model kink structures and the complete washing out of such kinks in the
PQMVT/QMVT model results, one has to investigate, analyze and compare the scalar and pseudo scalar meson mixing angles.
\subsection{Meson Mixing Angle Variations}
		We will finally be investigating the behavior of the scalar $\theta_S$ and pseudo scalar 
$\theta_P$ mixing angles. In the Fig.\ref{fig:mini:fig7:a}, the lower (upper) solid line in black color and
dash dot line in red color, depict the $\theta_P$ ($\theta_S$) variations respectively in the PQMVT and PQM
model computations for non zero axial anomaly. In the absence of axial anomaly, the dot like small dash line in
magenta color and the dash line in dark blue color, represent the respective scalar mixing angle $\theta_S$
variations for the PQMVT and PQM model. The same line types in Fig.\ref{fig:mini:fig7:b}, show the $\theta_P$
and $\theta_S$ variations for the QMVT and the QM model. The pseudo scalar $\theta_P$ mixing angle variations
shown by the filled green circles in both the figures, are constant when the axial anomaly is zero in all the
model calculations. Comparing the PQMVT(QMVT) model variations of $\theta_P$ and $\theta_S$ in 
Fig.\ref{fig:mini:fig7:a}(Fig.\ref{fig:mini:fig7:b}) with the corresponding PQM(QM) model results, we infer that
the fermionic vacuum correction  significantly modifies the axial $U_A(1)$ restoration pattern.
\begin{figure*}[!htbp]
\subfigure[With $U_A(1)$ anomaly term]{
\label{fig:mini:fig8:a} 
\begin{minipage}[b]{0.45\linewidth}
\centering \includegraphics[width=3.45in]{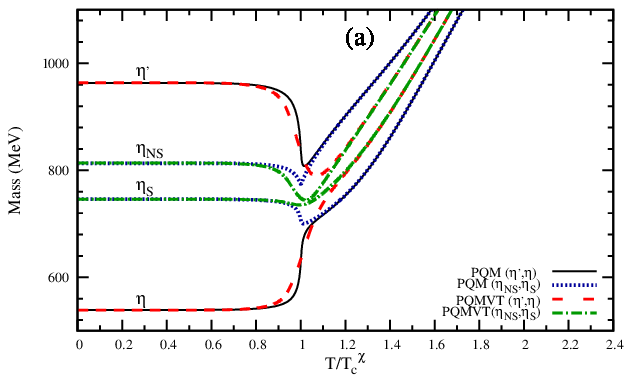}
\end{minipage}}%
\hspace{0.10in}
\subfigure[With $U_A(1)$ anomaly term]{
\label{fig:mini:fig8:b} 
\begin{minipage}[b]{0.45\linewidth}
\centering \includegraphics[width=3.45in]{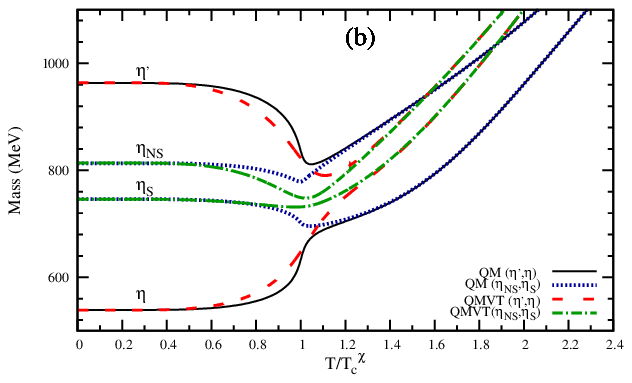}
\end{minipage}}
\caption{Shows the mass variations for the physical $\eta'$, $\eta$ and the 
non strange-strange $\eta_{NS}$, $\eta_{S}$ complex, on the reduced 
temperature scale ($T/T_{c}^{\chi}$) at zero chemical potential ($\mu = 0$).
Fig.\ref{fig:mini:fig8:a} shows the results for PQMVT and PQM model and line types
for mass variations are  labeled. Fig.\ref{fig:mini:fig8:b} shows the 
mass variations for the QMVT and QM model with labeled line types.}
\label{fig:mini:fig8} 
\end{figure*}
\begin{figure*}[!htbp]
\subfigure[With $U_A(1)$ anomaly term]{
\label{fig:mini:fig9:a} 
\begin{minipage}[b]{0.45\linewidth}
\centering \includegraphics[width=3.4in]{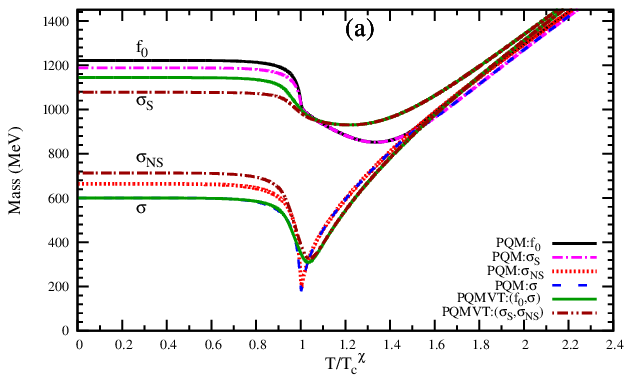}
\end{minipage}}%
\hspace{0.10in}
\subfigure[With $U_A(1)$ anomaly term]{
\label{fig:mini:fig9:b} 
\begin{minipage}[b]{0.45\linewidth}
\centering \includegraphics[width=3.4in]{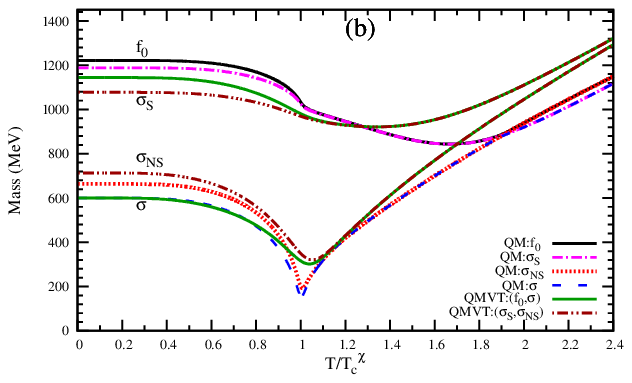}
\end{minipage}}
\caption{Shows the mass variations for the physical $\sigma$, $f_{0}$ and the 
non strange-strange $\sigma_{NS}$, $\sigma_{S}$ complex, on the reduced 
temperature scale ($T/T_{c}^{\chi}$) at zero chemical potential ($\mu = 0$).
Fig.\ref{fig:mini:fig9:a} shows the results for PQMVT and PQM model and line types
for mass variations are  labeled. Fig.\ref{fig:mini:fig9:b} shows the 
mass variations for the QMVT and QM model with labeled line types.
The masses of the physical $\sigma$ and $f_0$ anti-cross and the 
non strange-strange $\sigma_{NS} - \sigma_{S}$ system masses cross for 
the PQM/QM model variations. Such crossing and anti-crossing of masses disappears from 
the PQMVT/QMVT model results.}
\label{fig:mini:fig9} 
\end{figure*}

The non strange and strange quark mixing is strong as in Ref. \cite{Schaefer:09}, and one gets almost 
constant pseudo scalar mixing angle $\theta_P=-5^{\circ}$ in all the models when axial anomaly is present for the 
chiral symmetry broken phase at $T=0$. The $\theta_P$ variation near $T/T_c^{\chi}=1$
in the PQMVT model in Fig.\ref{fig:mini:fig7:a} develops a small dip and then smoothly starts the approach
toward the ideal mixing angle $\theta_P\to \arctan{\frac{1}{\sqrt{2}}}\sim 35^{\circ}$,
the corresponding  $\Phi_P = 90^\circ$. Here $\Phi_P$ is the pseudo scalar mixing angle in the 
strange non strange basis (see Ref.\cite{Schaefer:09} for details). In computations with the presence of
axial anomaly for $T/T_c^{\chi}>1$, the pseudo scalar mixing angle approaches  its ideal value  more smoothly
in the PQMVT(QMVT) model when compared with the corresponding result in the PQM(QM) model in Fig.\ref{fig:mini:fig7:a}
(Fig.\ref{fig:mini:fig7:b}). This approach is sharpest in the PQM model.

	The $\eta$ and $\eta'$ mesons become a purely strange $\eta_S$ 
and non strange $\eta_{NS}$ quark system as a consequence of the 
ideal pseudo scalar mixing which gets fully achieved at higher 
values of the reduced temperature. In order to show this and make comparisons,
the mass variations for the physical $\eta$, $\eta'$ and the non strange-strange $\eta_{NS}$, 
$\eta_S$ complex, are plotted for the PQMVT(QMVT) and PQM(QM) model in the Fig.\ref{fig:mini:fig8:a}(Fig.\ref{fig:mini:fig8:b}).
Mass formula $m_{\eta_{NS}}$ and $m_{\eta_S}$ are given in the Table \ref{tab:mass} of appendix B. 
In the $m_{\eta'}$ approach to $m_{\eta_{NS}}$ and the $m_{\eta}$ approach to $m_{\eta_S}$ around $T/T_c^{\chi} = 1$,
the most smooth and smoother mass convergence trend is seen respectively in the QMVT and PQMVT model in 
Fig.\ref{fig:mini:fig8:b} and Fig.\ref{fig:mini:fig8:a}.
Comparing the mass difference of $m_{\eta'}$  and $m_{\eta}$ for $T/T_c^{\chi} > 1$ in different models, the smallest
difference seen in the PQMVT model indicates that, here, we are getting the most converging
$U_A(1)$ restoration trend on account of the fermionic vacuum term .

	The fermionic vacuum correction gives rise to a decreased scalar mixing
angle $\theta_S$ in vacuum ($T=0$) . It becomes 11.98(14.75) degree in the presence(absence) 
of axial anomaly in the PQMVT and QMVT models  in Fig.\ref{fig:mini:fig7:a} and Fig.\ref{fig:mini:fig7:b} respectively from 
its value of 19.86(21.5) degree in the PQM and QM models. The $\theta_S$ growth to its ideal value near $T/T_{c}^{\chi}=1$, 
is smoother in the PQMVT and QMVT model. The most striking effect of the fermionic vacuum correction can be 
seen in the complete modification of the $\theta_S$  behavior in the higher temperature chirally symmetric phase 
of the PQMVT and QMVT model for both the cases with and without the axial anomaly. Instead of dropping down to the negative 
values as in the chiral symmetry restored phase of the  PQM and QM model respectively in the 
Fig.\ref{fig:mini:fig7:a} and Fig.\ref{fig:mini:fig7:b}, the $\theta_S$ in the PQMVT and QMVT model,  approaches the ideal
mixing angle very smoothly analogous to the pseudo scalar mixing angle  $\theta_P$ temperature variation computed in 
the presence of axial anomaly.

	In the chirally symmetric phase of the PQM and QM model respectively in the Fig.\ref{fig:mini:fig7:a} and Fig.\ref{fig:mini:fig7:b},
 the scalar mixing angle first achieves its ideal value $35^{\circ}$ and then drops down to $\theta_S \sim -51^{\circ}(-54^{\circ})$ for
higher temperatures in the presence (absence) of $U_A(1)$ axial symmetry breaking term. This drop happens around $T/T_c^{\chi} \sim 1.5(1.9)$ in 
the PQM(QM) model for non zero $c$ and similar drop for the calculations without the axial anomaly occurs at a 
little higher value of $T/T_c^{\chi}$. This pattern is already reported and discussed in 
Ref.\cite{Schaefer:09,gupta}. We note that when $\theta_S \sim 35^{\circ}$, the $f_0$ meson degenerates with 
the pure strange quark system $\sigma_S$ while $\sigma$ meson becomes identical with the pure non-strange quark system $\sigma_{NS}$.
Since the mixing angle $\sim -55^{\circ}$ is complimentary to $35^{\circ}$ (their difference is $\sim 90^{\circ}$), the temperature
variations of masses show a mass reversal trend in the non strange - strange basis  when $\theta_S \sim -51^{\circ}$ and the physical
$\sigma$ and $f_0$ masses anti-cross while the non strange - strange 
($\sigma_{NS} - \sigma_S$) system masses cross near the above mentioned reduced temperatures. After
anti-crossing the physical $\sigma$ becomes identical with pure 
strange quark system $\sigma_S$ while the physical $f_0$ 
degenerates with the pure non strange quark system $\sigma_{NS}$.
In order to show this crossing-anti crossing behavior in the presence of axial anomaly,
we have plotted in the Fig.\ref{fig:mini:fig9:a} and Fig.\ref{fig:mini:fig9:b}, the respective 
PQM and QM model mass variations for the physical $\sigma$ and $f_{0}$ and the 
non strange-strange $\sigma_{NS}$, $\sigma_S$ complex. 
Since the effect of fermionic vacuum fluctuation drastically 
modifies the $\theta_S$ behavior for higher temperatures, the masses of the physical
$\sigma$ and $f_0$ do not anti-cross and the non strange - strange 
($\sigma_{NS} - \sigma_S$) system masses do not cross for higher values on the 
reduced temperature scale and the $\sigma$ becomes identical with the pure 
non strange quark system $\sigma_{NS}$ while the physical $f_0$ smoothly 
degenerates with the pure strange quark system $\sigma_{S}$, in the PQMVT 
and QMVT model plots respectively in the Fig.\ref{fig:mini:fig9:a} and Fig.\ref{fig:mini:fig9:b}.
Here the $\theta_S$ approaches $\sim 35^{\circ}$ and then remains the same for higher temperatures.
\section{Summary and conclusion}
\label{sec:smry}
In the present  work, we have investigated how the inclusion of properly renormalized fermionic vacuum 
fluctuation in the 2+1 flavor QM and PQM models, modifies the  finite temperature behavior of
masses and mixing angles of scalar and pseudo scalar mesons. It  has been  explicitly 
shown that expressions for the  model parameters, meson masses and mixing angles, do not
depend on any arbitrary renormalization scale. We explored the qualitative and quantitative effects 
of fermionic vacuum correction, on the emerging mass degeneration patterns in the temperature variations
of masses of the chiral partners in pseudo scalar ($\pi$, $\eta$, $\eta'$, $K$) and scalar ($\sigma$, $a_0$, $f_0$,$\kappa$) 
meson nonets. From the mass convergence patterns, we identified chiral symmetry and $U_A(1)$ restoration
trends and compared them in different model scenarios. 

The fermionic vacuum correction causes a smoother and gentler crossover transition in the non-strange sector
where the transition temperature $T_{c}^{\chi}$ for the  PQMVT(QMVT) model increases by 10.7(25) MeV over its 
PQM(QM) model value. Unlike the PQM model result, the deconfinement crossover transition 
is not coincident with the non-strange sector chiral crossover transition for the PQMVT model calculation with
the $m_{\sigma}$=600, $T_0$=270 MeV and logarithmic ansatz for the Polyakov loop potential. However for
$m_{\sigma}$=400 MeV in the PQMVT model, we get $T_{c}^{\chi}$=202.6 MeV and $T_{c}^{\Phi}$=201.1 MeV. 
The sharpest PQM model $\sigma_x$ variation for the $T/T_{c}^{\chi}$ = 0.9 to 1.2 range  
becomes smoother and gentler in the PQMVT model. The QM model $\sigma_x$ temperature variation becomes a lot 
more smooth in the QMVT model only due to the effect of fermionic vacuum correction term.
The significant $\sigma_y$ melting of the PQM model gets further enhanced on account of the fermionic vacuum correction
and we obtain the largest but smoother melting of strange condensate  in the PQMVT model.

The sharpest mass degeneration of the PQM model for the ($\sigma$,$\pi$) and ($a_0$,$\eta'$) mesons, 
becomes quite smooth in the PQMVT model  and the most smooth mass degeneration results in the QMVT model.
We conclude from the behavior of these chiral partners
that the  chiral $SU_L(2) \times SU_R(2)$ symmetry restoring transition in the non strange sector, 
becomes quite smooth on account of the fermionic vacuum correction. In its influence,
the chiral symmetry restoration in the strange sector also becomes quite smooth and similar trends of smoother 
mass degeneration can be seen in the PQMVT/QMVT model temperature variations of the masses of the chiral 
partners ($\eta$,$f_0$) and ($K$, $\kappa$). It is worth emphasizing that the kink structure in
the PQM/QM model $m_\sigma$ and $m_{f_0}$ temperature variations altogether disappears from the corresponding
PQMVT/QMVT model results due to the noteworthy effect of the fermionic vacuum correction
and the $m_{f_0}$ degenerates quite smoothly with the $m_K$, $m_{\kappa}$ and $m_\eta$.
Further the smoothly merged line of $m_\sigma$ and $m_\pi$, shows a closer and narrower convergence to the degenerate 
line  of $m_{a_0}$ and $m_\eta'$ for higher $T/T_c^{\chi}>1$ in the PQMVT/QMVT model.
This behavior is a consequence of the largest but smoother melting of the $\sigma_y$
in the  PQMVT model because the $U_A(1)$ breaking anomaly effect that leads to the mass gap 
between the two sets of the chiral partners, ($\sigma$, $\pi$) and 
($a_0$, $\eta'$) i.e. $m_{\pi} = m_{\sigma}$ $<$ $m_{a_0}=m_{\eta'}$ 
for $T/T_C^{\chi}>1$, is proportional to the strange condensate 
$\sigma_y$. Thus the incorporation of fermionic vacuum correction 
in the PQM/QM  model, also effects an early set up of the $U_A(1)$ restoration trend 
on the reduced temperature scale. 

The pseudo scalar mixing angle $\theta_P$ in its approach to the ideal limit for $T/T_c^{\chi}>1$, also looks 
quite smoother in the PQMVT/QMVT model calculations with the axial anomaly. The smallest mass difference between
the $m_{\eta'}$(=$m_{\eta_{NS}}$) and $m_{\eta}$(=$m_{\eta_{S}}$) for $T/T_c^{\chi}>1$, results in the PQMVT model. 
It shows that the fermionic vacuum correction generates a most effective $U_A(1)$ restoration trend in the 
pseudo scalar sector. Further in its influence, the scalar mixing angle $\theta_{S}$ in the vacuum ($T=0$) 
decreases to 11.98(14.75) degree in the presence(absence) of axial anomaly in the PQMVT/QMVT model  from its value
of 19.86(21.5) degree for the PQM/QM model. In the chirally restored phase of the PQMVT/QMVT model, the fermionic vacuum
correction drastically modifies the PQM model behavior for scalar mixing angle which instead of becoming negative, approaches
its ideal value $\theta_{S} \sim 35^{\circ}$ quite smoothly for higher temperatures in both the presence as well as absence of
axial anomaly. As a consequence, unlike the PQM model, masses of the physical $\sigma$ and $f_0$ do not anti-cross and the 
non strange-strange ($\sigma_{NS}-\sigma_S$) system masses do not cross. The $\sigma$ becomes identical with the pure non strange quark
system $\sigma_{NS}$ while the physical $f_0$ smoothly degenerates with the pure strange quark system $\sigma_{S}$ and their mass 
variations become free of kink structure.
\begin{acknowledgments}
Valuable suggestions and computational helps given by Rajarshi Ray during the completion 
of this work are specially acknowledged. I am very much thankful to Rajarshi Tiwari for helping
me in generating good quality colored figures. General physics discussions with Ajit Mohan Srivastava are
very helpful. Computational support of the computing facility which has been developed by the Nuclear Particle
Physics group of the Physics Department, Allahabad University under the Center of Advanced Studies(CAS) 
funding of UGC, India, is also acknowledged. 
\end{acknowledgments}
\section{Appendix}
\label{sec.appendix}
\appendix
\section{Renormalized Model Parameters}
\label{modelparameters}
\begin{table}[!b]
\begin{tabular}{|cc|cccc|}
\hline
&& $m^{2}_{x,a} m^{2}_{x,b}/g^4$ & $m^{2}_{x,ab}/g^2$& $m^{2}_{y,a} m^{2}_{y,b}/g^4$ & $m^{2}_{y,ab}/g^2$\\
\hline
$\sigma_0$ & $\sigma_0$ &$\frac{1}{3}\sigma_{x}^{2}$& $\frac{2}{3}$& $\frac{1}{3}
\sigma_{y}^{2}$& $\frac{1}{3}$ \\
$\sigma_1$ & $\sigma_1$ &$\frac{1}{2}\sigma_{x}^{2}$& $ 1$ & $0$ & $0$\\
$\sigma_4$ & $\sigma_4$ &$ 0$ &$\sigma_x \frac{\sigma_x + \sqrt{2} \sigma_y}
{\sigma_{x}^{2} -2 \sigma_{y}^{2}}$ & $0$ & $\sigma_y \frac{ \sqrt{2} \sigma_x +2 \sigma_y}{2 \sigma_{y}^{2} - \sigma_{x}^{2}}$ \\
$\sigma_8$ & $\sigma_8$ &$ \frac{1}{6} \sigma_{x}^{2}$ & $\frac{1}{3}$ & 
$ \frac{2}{3}\sigma_{y}^{2} $ & $\frac{2}{3} $ \\
$\sigma_0$ & $\sigma_8$ & $ \frac{\sqrt{2} }{6} \sigma_{x}^{2}$ & $ \frac{\sqrt{2}}{3}$ & 
$- \frac{\sqrt{2}}{3} \sigma_{y}^{2}$ & $- \frac{\sqrt{2}}{3}$\\ 
$\pi_0$ & $\pi_0$ & $0$ & $\frac{2}{3}$ & $0$ & $\frac{1}{3}$ \\
$\pi_1$ & $\pi_1$ & $0$ & $1$ & $0$ & $0$ \\
$\pi_4$ & $\pi_4$ & $0$ & $\sigma_x \frac{\sigma_x -\sqrt{2}\sigma_y}{\sigma_{x}^{2} 
- 2 \sigma_{y}^{2}}$ & $0$ & $\sigma_y \frac{\sqrt{2} \sigma_x -2 \sigma_y}
{\sigma_{x}^{2} -2 \sigma_{y}^{2}}$ \\
$\pi_8$ & $\pi_8$ & $0$ & $\frac{1}{3}$ & $0$ & $\frac{2}{3}$\\
$\pi_0$ & $\pi_8$ & $0$ & $\frac{\sqrt{2}}{3}$ & $0$ & $-\frac{\sqrt{2}}{3}$ \\ 
\hline
 \end{tabular}
\caption{First and second derivative of squared quark mass in 
non strange-strange basis with respect to meson fields are evaluated 
at minimum. The symbol x in the first two columns denotes the sum over two light flavors.
The last two columns have only strange quark mass flavor denoted by symbol y.}
\label{tab:qmassd}
\end{table}
The mass modification in vacuum ($T=0$, $\mu=0$), due to the fermionic vacuum correction will be given by
\bea
&&({\delta m^{\text{v}}_{\alpha,ab}})^{2}=\frac{\partial^2 \Omega_{q\bar{q}}^{\rm vac}}
{\partial \xi_{\alpha,a}\partial \xi_{\alpha,b}} \bigg|_{min} \nn \\
&&=-\frac{N_c}{8\pi^2}\sum_f\Bigl[\l
2\log\l\frac{m_f}{\text{M}}\r+\frac{3}{2} \r\l\frac{\partial m_f^2}{\partial
 \xi_{\alpha,a}}\r\l\frac{\partial m_f^2}{\partial \xi_{\alpha,b}}\r \nn \\
&&+\l\frac{m_f^2}{2}+2m_f^2\log\l\frac{m_f}{\text{M}} \r \r \frac{\partial^2
  m_f^2}{\partial \xi_{\alpha,a}\partial \xi_{\alpha,b}}\Bigr]
\label{eq:dFVac}
\eea
\begin{table*}[!hbt]
\begin{tabular}{|l|l||l|l|}
 \hline
    & Meson masses calculated from pure mesonic potential  & & Fermionic vacuum correction in meson masses \\\hline
    $(m^{\text{m}}_{a_{0}})^2$ & $m^2 +\lambda_1(x^2+y^2)+\frac{3\lambda_2}{2} x^2+
    \frac{\sqrt{2}c}{2}y $\qquad & $({\delta m^{\text{v}}_{s,11}})^{2}$ & $-\frac{N_cg^4}{64\pi^2}x^2(4+3X)$
    \\ 
    $(m^{\text{m}}_{\kappa})^{2}$&$m^2+\lambda_1(x^2+y^2)+\frac{\lambda_{2}}{2}(x^2+\sqrt{2}xy+2y^2)+\frac{c}{2} x $\qquad & $({\delta m^{\text{v}}_{s,44}})^{2}$
    & $-\frac{N_cg^4}{64\pi^2}\l\frac{x+\sqrt{2}y}{x^2-2y^2}\r \l x^3X-2\sqrt{2}y^3Y\r$
    \\
    $(m^{\text{m}}_{s,00})^2$ &  $m^2+\frac{\lambda_1}{3}(7x^2+4\sqrt{2}xy+5y^2)+\lambda_2(x^2 + y^2)-\frac{\sqrt{2}c}{3} (\sqrt{2} x +y)$ &
    $({\delta m^{\text{v}}_{s,00}})^{2}$&$-\frac{N_cg^4}{96\pi^2}\l3\l x^2X+y^2Y\r+4\l x^2+y^2\r\r$ \\
    $(m^{\text{m}}_{s,88})^{2}$ & $m^2 +\frac{\lambda_1}{3}(5x^2-4\sqrt{2}xy +7y^2)+\lambda_2(\frac{x^2}{2} +2y^2)+\frac{\sqrt{2}c}{3} (\sqrt{2}x-\frac{y}{2})$ & 
    $({\delta m^{\text{v}}_{s,88}})^{2}$&  
    $-\frac{N_cg^4}{96\pi^2}\l \frac{3}{2}\l x^2X+4y^2Y\r+2\l x^2+4y^2\r\r$
    \\
   $(m^{\text{m}}_{s,08})^{2}$ &  $\frac{2\lambda_1}{3}(\sqrt{2}x^2 -xy -\sqrt{2}y^2) +\sqrt{2}\lambda_2(\frac{x^2}{2}-y^2) +\frac{c}{3\sqrt{2}}(x- \sqrt{2}y)$ &
  $({\delta m^{\text{v}}_{s,08}})^{2}$&$-\frac{N_cg^4}{8\sqrt{2}\pi^2}\l\frac{1}{4}\l x^2X-2y^2Y\r+\frac{1}{3}\l x^2-2y^2\r\r$   
   \\
$ (m^{\text{m}}_{\pi})^{2}$ & $m^2 + \lambda_1 (x^2 + y^2) +\frac{\lambda_2}{2} x^2 -\frac{\sqrt{2} c}{2} y$& $({\delta m^{\text{v}}_{p,11}})^{2}$  
&$-\frac{N_cg^4}{64\pi^2}x^2X$  \\ 
$(m^{\text{m}}_{K})^{2}$ &$m^2 + \lambda_1 (x^2 + y^2) +\frac{\lambda_2}{2} (x^2 - \sqrt{2} x y +2 y^2) - \frac{c}{2} x$&$({\delta m^{\text{v}}_{p,44}})^{2}$
&$-\frac{N_cg^4}{64\pi^2}\l\frac{x-\sqrt{2}y}{x^2-2y^2}\r \l x^3X+2\sqrt{2}y^3Y\r$  \\ 
$(m^{\text{m}}_{p,00})^{2}$ & $m^2 + \lambda_1(x^2 +y^2) + \frac{\lambda_2}{3}(x^2 +y^2) + \frac{c}{3} (2x + \sqrt{2} y)$&$({\delta m^{\text{v}}_{p,00}})^{2}$
&$-\frac{N_cg^4}{96\pi^2}\l x^2X+y^2Y\r$ \\
$(m^{\text{m}}_{p,88})^{2}$ & $m^2 +\lambda_1(x^2 +y^2) +\frac{\lambda_2}{6}(x^2 +4y^2)-\frac{c}{6}(4x -\sqrt{2}y)$& $({\delta m^{\text{v}}_{p,88}})^{2}$
&$-\frac{N_cg^4}{192\pi^2}\l x^2X+4y^2Y\r$ \\
$(m^{\text{m}}_{p,08})^{2}$ & $\frac{\sqrt{2}\lambda_2}{6}(x^2-2y^2)-\frac{c}{6}(\sqrt{2}x -2y)$& $({\delta m^{\text{v}}_{p,08}})^{2}$
&$-\frac{N_cg^4}{96\sqrt{2}\pi^2}\l x^2X-2y^2Y\r$ \\ 
\hline
\end{tabular}
\caption{Expressions for $({m^{\text{m}}_{\alpha,ab}})^{2}$ in the left half of the table are the vacuum
meson masses calculated from the second derivatives of the pure mesonic potential U($\sigma_x,\sigma_y$).
Evaluated expressions of mass modifications $({\delta m^{\text{v}}_{\alpha,ab}})^{2}$ due to the 
fermionic vacuum correction are given in the right half. Symbols used in the expressions are defined as $x=\sigma_x$, 
$y=\sigma_y$, $X=\l1+4\log\l \frac{g\sigma_x}{2\text{M}}\r\r$ and $Y=\l1+4\log\l \frac{g\sigma_y}{\sqrt{2}\text{M}}\r\r$.}
\label{tab*:MesonVM}
\end{table*}
Here $|_{min}$ stands for the global minimum of the full grand potential in Eq.(\ref{eq:grandp}).
The first $m^{2}_{f,a} \equiv \partial m^{2}_{f}/ \partial \xi_{\alpha,a}$ and second  
$m^{2}_{f,ab} \equiv \partial m^{2}_{f,a}/ \partial \xi_{\alpha,b}$ partial derivatives of squared quark mass
with respect to the meson fields as evaluated in Ref. \cite{Schaefer:09} in the non strange-strange basis, 
are presented in the Table \ref{tab:qmassd}. We have evaluated the mass modifications given in Eq.(\ref{eq:dFVac})
and collected different expressions of $({\delta m^{\text{v}}_{\alpha,ab}})^{2}$ in the Table \ref{tab*:MesonVM}
for all the mesons of scalar and pseudo-scalar nonet. The vacuum mass expressions $({m^{\text{m}}_{\alpha,ab}})^{2}$
for these mesons as originally evaluated from the second derivative of pure mesonic potential in 
Ref.~\cite{Rischke:00,Schaefer:09}, are also given in this Table. Here, when $\alpha$ =s, the (11)
element gives squared mass of the scalar $a_{0}$ meson which is degenerate with the (22) and (33) elements.
Similarly, the (44) element which is degenerate with (55), (66) and (77) elements, gives 
the squared $\kappa$ meson mass. The squared  $\sigma$ and $f_{0}$ meson masses are obtained by diagonalizing the scalar mass matrix in
the (00)-(88) sector and we get a scalar mixing angle $\theta_{S}$. We have a completely analogous situation for the pseudo scalar sector
($\alpha$ = p)  with the following identification; the (11) element gives the squared pion mass and the squared kaon mass
is given by the (44) element. Further the diagonalization of the (00)-(88) sector of the pseudo scalar mass matrix, gives 
the squared masses of $\eta$ and  $\eta'$ and analogously, we get a pseudo-scalar mixing angle $\theta_{P}$. 
\begin{widetext}
In the QMVT/PQMVT model calculations, the vacuum mass expressions in Eq.(\ref{eq:derVac}) that determine 
$\lambda_2$ and c are $m^{2}_{\pi}=({m^{\text{m}}_{\pi}})^{2}+({\delta m^{\text{v}}_{p,11}})^{2}$, 
$m^{2}_K=({m^{\text{m}}_K})^{2}+({\delta m^{\text{v}}_{p,44}})^{2}$ and $m^{2}_{\eta}+m^{2}_{\eta'}=m^{2}_{p,00}+m^{2}_{p,88}$ where
$m^{2}_{p,00}=({m^{\text{m}}_{p,00}})^{2}+({\delta m^{\text{v}}_{p,00}})^{2}$ and 
$m^{2}_{p,88}=({m^{\text{m}}_{p,88}})^{2}+({\delta m^{\text{v}}_{p,88}})^{2}$. We can write $m^{2}_{\eta}+m^{2}_{\eta'}=({m^{\text{m}}_{\eta}})^{2}+({m^{\text{m}}_{\eta'}})^{2}+({\delta m^{\text{v}}_{p,00}})^{2}+({\delta m^{\text{v}}_{p,88}})^{2}$ where $({m^{\text{m}}_{\eta}})^{2}+({m^{\text{m}}_{\eta'}})^{2}=
({m^{\text{m}}_{p,00}})^{2}+({m^{\text{m}}_{p,88}})^{2}$. Using mass modification  expressions
$({\delta m^{\text{v}}_{\alpha,ab}})^{2}$  given in the Table \ref{tab*:MesonVM}, we write
\bea
&&({m^{\text{m}}_K})^{2}=m^{2}_K+\frac{N_cg^4}{64\pi^2}\l\frac{x-\sqrt{2}y}{x^2-2y^2}\r 
\l x^3X+2\sqrt{2}y^3Y\r\ ; \ ({m^{\text{m}}_{\pi}})^{2}=m^{2}_{\pi}+
\frac{N_cg^4}{64\pi^2}x^2X\  \text{and}  \nn\\ 
&&\l({m^{\text{m}}_{\eta}})^{2}+({m^{\text{m}}_{\eta'}})^{2}\r=\l m^{2}_{\eta}+m^{2}_{\eta'}\r+
\frac{N_cg^4}{192\pi^2}\l3x^2X+6y^2Y\r\
\label{PIKETA}
\eea
The $f_{\pi}$ and $f_K$ give vacuum condensates according to the partially conserved axial vector current (PCAC) relation. The
x =$\sigma_x=f_{\pi}$ and y =$\sigma_y=\l \frac{2f_K-f_{\pi}}{\sqrt{2}}\r$ at $T=0$. The parameters $\lambda_2$ and c  
in vacuum are obtained as:
\bea
\lambda_2&=&\frac{3\l2f_K-f_{\pi}\r({m^{\text{m}}_K})^{2}-\l2f_K+f_{\pi}\r({m^{\text{m}}_{\pi}})^{2}
-2\l({m^{\text{m}}_{\eta}})^{2}+({m^{\text{m}}_{\eta'}})^{2}\r 
\l f_K-f_{\pi}\r}{\l3f^2_{\pi}+8f_K\l f_K-f_{\pi}\r\r\l f_K-f_{\pi}\r}\ 
\label{lam2c}
\eea
\be
c=\frac{({m^{\text{m}}_K})^{2}-({m^{\text{m}}_{\pi}})^{2}}{f_K-f_{\pi}}-\lambda_2\l2f_K-f_{\pi}\r
\label{eqc}
\ee
When expressions of $({m^{\text{m}}_{\pi}})^2$, $({m^{\text{m}}_K})^2$ and $\l ({m^{\text{m}}_{\eta}})^2+
({m^{\text{m}}_{\eta'}})^2\r$ from Eq.(\ref{PIKETA}) are substituted in Eq.(\ref{lam2c}) and Eq.(\ref{eqc}) 
and the vacuum value of the condensates are used, the final rearrangement of terms yields:
\bea
\label{lam2VT}
\lambda_2&=&\lambda_{2s}+n+\lambda_{2+}+\lambda_{\text{2M}}\ \text{where} \ \lambda_{2s}=\frac{3\l2f_K-f_{\pi}\r m^{2}_K-\l2f_K+f_{\pi}\r m^{2}_{\pi}-2\l m^{2}_{\eta}+m^{2}_{\eta'}\r \l f_K-f_{\pi}\r}{\l3f^2_{\pi}+8f_K\l f_K-f_{\pi}\r\r\l f_K-f_{\pi}\r}\ ,\nn\\
n&=&\frac{N_cg^4}{32\pi^2}, 
\ \lambda_{2+}=\frac{n{f_{\pi}}^2}{f_K \l f_K-f_{\pi}\r}\log\l 
\frac{2\ f_K-f_{\pi}}{f_{\pi}}\r
\text{and scale dependent part} \ \lambda_{\text{2M}}= 4n\log\l \frac{g\l 2f_K-f_{\pi}\r}{2 \text{M}}\r
\eea
\be 
\label{cfinal}
c=\frac{{m_K}^{2}-{m_{\pi}}^{2}}{f_K-f_{\pi}}-\lambda_{2s}\l2f_K-f_{\pi}\r
\ee
\end{widetext}
We note that the $\lambda_{2s}$ is equal to the earlier
$\lambda_2$ parameter determined in the QM/PQM model calculations in 
Ref.~\cite{Rischke:00,gupta,Schaefer:09}. In the present calculation, the proper renormalization 
of fermionic vacuum, leads to the augmentation of $\lambda_{2s}$ by the
addition of a term (n+$\lambda_{2+}$) and further, we get a renormalization scale M dependent 
contribution $\lambda_{\text{2M}}$ in the expression of the $\lambda_2$ in Eq.(\ref{lam2VT}).

We get the complete cancellation of M dependence in the evaluation of $c$ also and finally 
its value turns out to be the same as in the QM model. 
The scale M independent expression of $m^2_{\pi}$ obtained in the appendix B can be used with 
x=$f_{\pi}$ and y=$(\frac{2f_K-f_{\pi}}{\sqrt{2}})$, to express $m^2$ in terms of $\lambda_1$.
\bea
m^2&=&m^{2}_{\pi}-\lambda_1\{f_{\pi}^2+\frac{(2f_K-f_{\pi})^2}{2}\}-\frac{f_{\pi}^2}{2} \ \biggl[\ \lambda_{2\text{v}}-\ 4 \ n\nn\\
&& \log \ \{\ \frac{f_{\pi}}{\l2f_K-f_{\pi}\r} \}\biggl] +\frac{c}{2} \ (\ 2f_K-f_{\pi})
\eea
When we use the formula of $m^2_{\sigma}$ in the Table \ref{tab:mass} of the appendix B (with the vacuum values of the masses $m^2_{s,00}$, $m^2_{s,88}$, $m^2_{s,08}$ and the mixing angle $\theta_S$) and 
substitute the above expression of $m^2$ in it, we will get the numerical value of $\lambda_1$ for 
$m_\sigma$=600 MeV and we will put $m_\pi$=138 MeV . The explicit symmetry breaking parameters $h_x$ 
and $h_y $ do not change due to the fermionic vacuum correction. 
\section{ Scale Independent Meson Masses}
\label{ScaleI:Masses}
When the value of $\lambda_2$ in the appendix A is substituted in the 
mass expressions $({m^{\text{m}}_{\alpha,ab}})^{2}$, the
logarithmic M dependence of $\lambda_2$ neatly cancels with the scale dependence already existing
in the mass modifications $({\delta m^{\text{v}}_{\alpha,ab}})^{2}$ due to the fermionic vacuum 
correction and the final expressions of meson masses $m^2_{\alpha,ab}$ become free of any 
scale dependence when these two contributions are added together.The mixing angles obtained
from these masses will naturally be independent of renormalization scale. The  expressions of
the scale independent meson masses are derived in the following. 
\begin{widetext}
Substituting the value of $\lambda_2$ from Eq.(\ref{lam2VT}) in the respective terms $(m^{\text{m}}_{a_{0}})^{2}$, $({m^{\text{m}}_\kappa})^{2}$, $(m^{\text{m}}_{s,00})^2$, $(m^{\text{m}}_{s,88})^2$ and $(m^{\text{m}}_{s,08})^2$ of the corresponding formulae
$m^{2}_{a_{0}}=(m^{\text{m}}_{a_{0}})^{2}+({\delta m^{\text{v}}_{s,11}})^{2}$, $m^{2}_\kappa=({m^{\text{m}}_\kappa})^{2}+({\delta m^{\text{v}}_{s,44}})^{2}$, $m^2_{s,00}=(m^{\text{m}}_{s,00})^2+({\delta m^{\text{v}}_{s,00}})^{2}$, $m^2_{s,88}=(m^{\text{m}}_{s,88})^2+({\delta m^{\text{v}}_{s,88}})^{2}$ and $m^2_{s,08}=(m^{\text{m}}_{s,08})^2+({\delta m^{\text{v}}_{s,08}})^{2}$, we obtain the renormalization scale M independent mass formulae of all the  mesons in the scalar nonet. We write $\lambda_{2\text{v}}$ for $\lambda_{2s}+\lambda_{2+}$.
\bea
m^{2}_{a_{0}}&=&m^2 + \lambda_1 (x^2 + y^2) +\biggl[\lambda_{2s}+n+\lambda_{2+}+4n\log\{ \frac {g( 2f_K-f_{\pi})}{2\text{M}}\}\biggl] \frac {3 x^2}{2} +\frac{\sqrt{2} c}{2} y-n \biggl[4+3 \{1+4\log\l \frac{g x}{2\text{M}}\r\}\biggl]\ \frac {x^2} {2} \nn\\
&=&m^2+\lambda_1(x^2+y^2)+\biggl[\lambda_{2\text{v}}-4n\log\{ \frac{x}{\l2f_K-f_{\pi}\r}\}-\frac{4n}{3}\biggl] \frac{3 x^2} {2}+\frac{\sqrt{2}c}{2}y \eea
\bea
m^{2}_{\kappa}&=&m^2 + \ \lambda_1 \ ( \ x^2 + y^2\ ) +\biggl[\lambda_{2s} \ + \ n+ \ \lambda_{2+} \ +\ 4\ n \ \log\ \{ \ \frac{g \ \l 2 \ f_K \ -\ f_{\pi}\r}{2\text{M}} \ \}\biggl] \frac{(\ x^2 +\sqrt{2} \ x \ y +  2 \ y^2 \ )}{2} \ +  \frac{c}{2} \ x \nn\\
&&-\frac{n}{2}\l\frac{x+\sqrt{2} y}{x^2-2 y^2}\r \biggl[ x^3 \{1+4\log\l \frac{g x}{2\text{M}}\r\}-2\sqrt{2}y^3 \{1+4\log\l \frac{g y}{\sqrt{2}\text{M}}\r\}\biggl]\  \nn\\
&=&m^2+\lambda_1(x^2+y^2)+\biggl[\lambda_{2\text{v}}-4n\log\{ \frac{x}{\l2f_K-f_{\pi}\r}\}\biggl] \frac{(x^2+\sqrt{2}xy +2y^2)}{2}+\frac{c}{2}x+\frac{4\sqrt 2 \ n y^3}{\l x-\sqrt 2 y\r}\log \frac{\sqrt 2 y}{x}
\eea
\bea
m^2_{s,00}&=&m^2+\frac{\lambda_1}{3}(7x^2+4\sqrt{2}xy+5y^2)+\biggl[\lambda_{2s}+n+\lambda_{2+}+4\ n\ \log \ \{\ \frac{g \ \l 2\ f_K-f_{\pi}\r}{2\text{M}}\}\biggl](x^2 + y^2)-\frac{\sqrt{2} \ c}{3} (\sqrt{2} \ x +y) \nn\\
&&-\frac{n}{3}\biggl[3\l x^2 \{1+4\log \frac{g x}{2\text{M}}\} +y^2\{1+4\log \frac{g y}{\sqrt{2}\text{M}} \}\r+4\l x^2+y^2\r\biggl] \nn\\
& =&m^2+\frac{\lambda_1}{3} (7x^2+4\sqrt{2}xy+5y^2)+(\lambda_{2\text{v}}-\frac{4n}{3}) (x^2+y^2)-4n\biggl[x^2 \log \frac{x}{\l2f_K-f_{\pi}\r}+y^2 \log \frac{\sqrt 2 y}{\l2f_K-f_{\pi}\r}\biggl] \nn\\
&&-\frac{c(2x +\sqrt{2}y)}{3} 
\eea 
\bea
m^2_{s,88}&=&m^2+\frac{\lambda_1}{3}(5x^2-4\sqrt{2}xy+7y^2)+\biggl[\lambda_{2s}+n+\lambda_{2+}+4n\log\{ \frac {g\l 2f_K-f_{\pi}\r}{2\text{M}}\}\biggl](\frac{x^2}{2} + 2 y^2)+\frac{\sqrt{2}c}{3} (\sqrt{2} x-\frac{y}{2}) \nn\\
&&-\frac{n}{6}\biggl[3\l x^2 \{1+4\log \frac{g x}{2\text{M}}\} +4y^2\{1+4\log \frac{g y}{\sqrt{2}\text{M}} \}\r+4\l x^2+4y^2\r\biggl] \nn\\
&=&m^2+\frac{\lambda_1}{3} (5x^2-4\sqrt{2}xy+7y^2)+(\lambda_{2\text{v}}-\frac{4n}{3}) (\frac{x^2}{2}+2y^2)-2n\biggl[x^2 \log \frac{x}{\l2f_K-f_{\pi}\r}+4y^2 \log \frac{\sqrt 2 y}{\l2f_K-f_{\pi}\r}\biggl] \nn\\
&&+\frac{c(2x -\frac{y}{\sqrt{2}})}{3}
\eea
\bea
m^2_{s,08}&=&2\frac{\lambda_1}{3}(\sqrt{2}x^2-xy-\sqrt{2}y^2)+\sqrt{2}\biggl[\lambda_{2s}+n+\lambda_{2+}+4n\log\{ \frac {g\l 2f_K-f_{\pi}\r}{2\text{M}}\}\biggl](\frac{x^2}{2} - y^2)+\frac{c}{3\sqrt{2}} (x-\sqrt{2}y) \nn\\
&&-\frac{n}{\sqrt{2}}\biggl[ x^2 \{1+4\log \frac{g x}{2\text{M}}\} -2y^2\{1+4\log \frac{g y}{\sqrt{2}\text{M}} \}+\frac{4}{3}\l x^2-2y^2\r\biggl] \nn\\
&=&2\frac{\lambda_1}{3}(\sqrt{2}x^2-xy-\sqrt{2}y^2)+\sqrt{2}(\lambda_{2\text{v}}-\frac{4n}{3}) (\frac{x^2}{2}-y^2)-2\sqrt{2}n\biggl[x^2 \log \frac{x}{\l2f_K-f_{\pi}\r}-2y^2 \log \frac{\sqrt 2 y}{\l2f_K-f_{\pi}\r}\biggl] \nn\\ 
&&+\frac{c(x -\sqrt{2}y)}{3\sqrt{2}}
\eea
Substituting the value of $\lambda_2$ from Eq.(\ref{lam2VT}) in the respective expressions of $({m^{\text{m}}_{\pi}})^{2}$,
$({m^{\text{m}}_K})^{2}$, $(m^{\text{m}}_{p,00})^2$, $(m^{\text{m}}_{p,88})^2$ and $(m^{\text{m}}_{p,08})^2$ in the corresponding formulae $m^{2}_{\pi}=({m^{\text{m}}_{\pi}})^{2}+({\delta m^{\text{v}}_{p,11}})^{2}$, 
$m^{2}_K=({m^{\text{m}}_K})^{2}+({\delta m^{\text{v}}_{p,44}})^{2}$, $m^2_{p,00}=(m^{\text{m}}_{p,00})^2+({\delta m^{\text{v}}_{p,00}})^{2}$,  $m^2_{p,88}=(m^{\text{m}}_{p,88})^2+({\delta m^{\text{v}}_{p,88}})^{2}$ and $m^2_{p,08}=(m^{\text{m}}_{p,08})^2+({\delta m^{\text{v}}_{p,08}})^{2}$, we obtain the following renormalization scale M independent mass formulae for the pseudo-scalar mesons :
\bea
m^{2}_{\pi}&=&m^2 + \lambda_1 (x^2 + y^2) +\biggl[\lambda_{2s}+n+\lambda_{2+}+4n\log\{ \frac {g( 2f_K-f_{\pi})}{2\text{M}}\}\biggl] \frac {x^2}{2} -\frac{\sqrt{2} c}{2} y-n \biggl[1+4\log\l \frac{g x}{2\text{M}}\r\biggl]\ \frac {x^2} {2} \nn\\
&=&m^2+\lambda_1(x^2+y^2)+\biggl[\lambda_{2\text{v}}-4n\log\{ \frac{x}{\l2f_K-f_{\pi}\r}\}\biggl] \frac{x^2} {2}-\frac{\sqrt{2}c}{2}y
\eea
\bea
m^{2}_K&=&m^2 + \lambda_1 (x^2 + y^2) +\biggl[\lambda_{2s}+n+\lambda_{2+}+4n\log\{ \frac {g\l 2f_K-f_{\pi}\r}{2\text{M}}\}\biggl] \frac {(x^2 - \sqrt{2} x y +2 y^2)}{2} -\frac{c}{2} x \nn\\
&&-\frac{n}{2}\l\frac{x-\sqrt{2} y}{x^2-2 y^2}\r \biggl[ x^3 \{1+4\log\l \frac{g x}{2\text{M}}\r\}+2\sqrt{2}y^3 \{1+4\log\l \frac{g y}{\sqrt{2}\text{M}}\r\}\biggl]\  \nn\\
&=&m^2+\lambda_1(x^2+y^2)+\biggl[\lambda_{2\text{v}}-4n\log\{ \frac{x}{\l2f_K-f_{\pi}\r}\}\biggl] \frac{(x^2-\sqrt{2}xy +2y^2)}{2}-\frac{c}{2}x-\frac{4\sqrt 2 \ n y^3}{\l x+\sqrt 2 y\r}\log \frac{\sqrt 2 y}{x}
\eea
\bea
m^2_{p,00}&=&m^2+\lambda_1 (x^2+y^2)+\biggl[\lambda_{2s}+n+\lambda_{2+}+4n\log\{ \frac {g\l 2f_K-f_{\pi}\r}{2\text{M}}\}\biggl]\frac{(x^2 + y^2)}{3}+\frac{\sqrt{2}c}{3} (\sqrt{2} x +y) \nn\\
&&-\frac{n}{3}\biggl[ x^2 \l1+4\log \frac{g x}{2\text{M}}\r +y^2\l1+4\log \frac{g y}{\sqrt{2}\text{M}} \r\biggl] \nn\\
&=&m^2+ \lambda_1 (x^2+y^2)+\frac{\lambda_{2\text{v}}}{3} (x^2+y^2)-\frac{4n}{3}\biggl[x^2 \log \frac{x}{\l2f_K-f_{\pi}\r}+y^2 \log \frac{\sqrt 2 y}{\l2f_K-f_{\pi}\r}\biggl]+\frac{c(2x +\sqrt{2}y)}{3} 
\eea 
\bea
m^2_{p,88}&=&m^2+\lambda_1(x^2+y^2)+\biggl[\lambda_{2s}+n+\lambda_{2+}+4n\log\{ \frac {g\l 2f_K-f_{\pi}\r}{2\text{M}}\}\biggl] \frac{(x^2+4y^2)}{6}-\frac{\sqrt{2}c}{3} (\sqrt{2} x-\frac{y}{2}) \nn\\
&&-\frac{n}{6}\biggl[ x^2 \l1+4\log \frac{g x}{2\text{M}}\r +4y^2\l1+4\log \frac{g y}{\sqrt{2}\text{M}} \r\biggl] \nn\\
& =&m^2+\lambda_1 (x^2+y^2)+\frac{\lambda_{2\text{v}}}{6} (x^2+4y^2)-\frac{2n}{3}\biggl[x^2 \log \frac{x}{\l2f_K-f_{\pi}\r}+4y^2 \log \frac{\sqrt{2} y}{\l2f_K-f_{\pi}\r}\biggl]-\frac{c(2x -\frac{y}{\sqrt{2}})}{3} 
\eea
\bea
m^2_{p,08}&=&\frac{\sqrt{2}}{6}\biggl[\lambda_{2s}+n+\lambda_{2+}+4n\log\{ \frac {g\l 2f_K-f_{\pi}\r}{2\text{M}}\}\biggl](x^2-2y^2)-\frac{c}{6} (\sqrt{2}x-2y)-\frac{n}{3\sqrt{2}}\biggl[ x^2 \l 1+4\log \frac{g x}{2\text{M}}\r \nn\\
&&-2y^2\l1+4\log \frac{g y}{\sqrt{2}\text{M}} \r\biggl]  \nn\\
& =& \frac{\sqrt{2} \lambda_{2\text{v}}}{6} (x^2-2y^2) -\frac{2\sqrt{2}n}{3}\biggl[x^2 \log \frac{x}{\l2f_K-f_{\pi}\r}-2y^2 \log \frac{\sqrt 2 y}{\l2f_K-f_{\pi}\r}\biggl]-\frac{c(\sqrt{2}x -2y)}{6} 
\eea
\begin{table*}[!hbt]
  \begin{tabular}{|l|l||l|l|} 
    \hline
    & Scalar Meson Masses & & Pseudo scalar Meson Masses \\\hline
    $m^{2}_{\sigma}$ & $m^2_{s,{00}}\cos^{2}\theta_{s}+m^2_{s,88}\sin^{2}\theta_{s}+2m^{2}_{s,08}\sin\theta_{s}\cos\theta_{s}$
    & $m^{2}_{\eta'}$ & $m^2_{p,{00}}\cos^{2}\theta_{p}+m^2_{p,88}\sin^{2}\theta_{p}+2m^{2}_{p,08}\sin\theta_{p}\cos\theta_{p}$ \\
    $m^{2}_{f_0}$ & $m^2_{s,00}\sin^{2}\theta_{s}+m^{2}_{s,88}\cos^{2}\theta_{s}-2m^2_{s,08}\sin\theta_{s}\cos\theta_{s}$\qquad
    & $m^{2}_{\eta}$ & $m^2_{p,00}\sin^{2}\theta_{p}+m^{2}_{p,88}\cos^{2}\theta_{p}-2m^2_{p,08}\sin\theta_{p}\cos\theta_{p}$\\
    $m^{2}_{\sigma_{NS}}$ & $\frac{1}{3}(2m^{2}_{s,00}+m^{2}_{s,88}+2\sqrt{2}m^{2}_{s,08})$ &
    $m^{2}_{\eta_{NS}}$ & $\frac{1}{3}(2m^{2}_{p,00}+m^{2}_{p,88}+2\sqrt{2}m^{2}_{p,08})$\\
    $m^{2}_{\sigma_{S}}$ & $\frac{1}{3}(m^{2}_{s,00}+2m^{2}_{s,88}-2\sqrt{2}m^{2}_{s,08})$ &
    $m^{2}_{\eta_{S}}$ & $\frac{1}{3}(m^{2}_{p,00}+2m^{2}_{p,88}-2\sqrt{2}m^{2}_{p,08})$\\
    \hline
  \end{tabular}
 \caption{The squared masses of scalar and pseudo scalar mesons which are obtained after the diagonalization of the 00-88 sector of
  mass matrix. The meson masses in the non strange $\sigma_{NS}(\eta_{NS})$ and strange $\sigma_S(\eta_S)$ basis are given in the
  last two rows.}
  \label{tab:mass}
\end{table*}
\end{widetext}


\begin{thebibliography}{199}
\bibitem{Shuryak}
E.V.Shuryak
Phys. Rep. {\bf61}, 71 (1980); ibid {\bf115}, 151 (1984).

\bibitem{Rafelski}
J. Rafelski
Phys. Rep. {\bf88}, 331 (1982); ibid {\bf142}, 167-262 (1986).

\bibitem{Svetitsky} 
L.D.McLerran, B.Svetitsky,
Phys. Rev. D{\bf24}, 450 (1981); 
B.Svetitsky, 
Phys. Rep. {\bf132}, 1 (1986).

\bibitem{Muller} 
B.Muller, 
Rep. Prog. Phys. {\bf 58}, 611 (1995).  

\bibitem{Ortmanns:96ea}
H.Meyer-Ortmanns
Rev.\ Mod.\ Phys.\ {\bf68}, 473 (1996).

\bibitem{Rischke:03}
D.~H.~Rischke,
  Prog. Part. Nucl. Phys. {\bf 52}, 197 (2004).

\bibitem{Polyakov:78plb}
A. M. Polyakov,
Phys. Lett. {\bf B 72}, 477 (1978).


\bibitem{Pisarski:00prd}
R. D. Pisarski,
Phys. Rev. {\bf D 62} 111501(R) (2000).


\bibitem{Vkt:06}
B. Layek, A. P. Mishra, A. M. Srivastava and V. K. Tiwari, 
Phys. Rev. {\bf D 73} 103514 (2006).

\bibitem{Kaczmarek:02}
O. Kaczmarek, F. Karsch, P. Petreczky and F. Zantow,
Phys. Lett. {\bf B 543}, 41 (2002).

\bibitem{Karsch:02}
F. Karsch,
Lect. Notes Phys. {\bf 583}, 209 (2002).

\bibitem{Fodor:03}
Z. Fodor, S. D. Katz, and K. K. Szabo,
Phys. Lett. {\bf B 568}, 73 (2003).

\bibitem{Allton:05}
C. R. Allton, M. Doring, S. Ejiri, S. J. Hands, O. Kaczmarek, F. Karsch, 
E Laermann and K. Redlich,
Phys. Rev. {\bf D 71}, 054508 (2005).

\bibitem{Karsch:05}
F. Karsch,
J. Phys. {\bf G 31}, S633 (2005).  

\bibitem{Karsch:07ax}
F. Karsch,
J. Phys. {\bf G 34}, S627 (2007); 
e-Print: arXiv:0701210 [hep-ph].


\bibitem{Aoki:06}
Y. Aoki, Z. Fodor, S. D. Katz and K. K. Szabo,
Phys. Lett. {\bf B 643}, 46 (2006).

\bibitem{FodorN}
Y.~Aoki, G.~Endrodi, Z.~Fodor, S.~D.~Katz and K.~K.~Szabo,
Nature {\bf 443}, 675 (2006) 


\bibitem{Fodor1}
S.~Borsanyi, G.~Endrodi, Z.~Fodor, A.~Jakovac, S.~D.~Katz, S.~Krieg, C.~Ratti and K.~K.~Szabo,
arXiv:1011.4229 [hep-lat].

\bibitem{HotQCD}
C.~Schmidt (HotQCD Collaboration)
AIP Conf.\ Proc.\ No. {\bf 1343}, 513 (AIP, New York, 2011).
arXiv:1012.2230 [hep-lat].

\bibitem{ChengLQCD}
M.~Cheng, N.~H.~Christ, S.~Datta, J.~van der Heide, C.~Jung, F.~Karsch, O.~Kaczmarek and E.~Laermann {\it et al.},
Phys.\ Rev.\ D {\bf 77}, 014511 (2008)

\bibitem{BazLQCD}
A.~Bazavov, T.~Bhattacharya, M.~Cheng, N.~H.~Christ, C.~DeTar, S.~Ejiri, S.~Gottlieb and R.~Gupta {\it et al.},
Phys.\ Rev.\ D {\bf 80}, 014504 (2009) 

\bibitem{WBLQCD}
S.~Borsanyi, G.~Endrodi, Z.~Fodor, A.~Jakovac, S.~D.~Katz, S.~Krieg, C.~Ratti and K.~K.~Szabo,
JHEP {\bf 1011}, 077 (2010)

\bibitem{LQCDWB2}
S.~Borsanyi, Z.~Fodor, C.~Hoelbling, S.~D.~Katz, S.~Krieg, C.~Ratti and K.~K.~Szabo,
JHEP {\bf 1009}, 073 (2010) ;  arXiv:1005.3508 [hep-lat].

\bibitem{LQCDWBL}
S.~Borsanyi, G.~Endrodi, Z.~Fodor, C.~Hoelbling, S.~D.~Katz, S.~Krieg, C.~Ratti and K.~K.~Szabo,
Acta Phys. Polon. Supp. {\bf 4}, 593-602 (2011) ; arXiv:1109.5032 [hep-lat].

\bibitem{HotLQCDL}
A.~Bazavov, T.~Bhattacharya, M.~Cheng, C.~DeTar, H.~T.~Ding, S.~Gottlieb, R.~Gupta and P.~Hegde {\it et al.},
Phys. Rev. {\bf D 85}, 054503 (2012); arXiv:1111.1710 [hep-lat].

\bibitem{Wilczek}
R. D. Pisarski and F. Wilczek
\newblock {Phys. Rev.} {\bf D 29}, 338 (1984).
 
\bibitem{Hatsuda:98}
S. Chiku and T. Hatsuda,
Phys. Rev. {\bf D 58}, 076001 (1998).

\bibitem{Bielich:00prl}
J. Schaffner-Bielich,
Phys. Rev. Lett. {\bf 84}, 3261 (2000).

\bibitem{Rischke:00}
J. T. Lenaghan, D. H. Rischke and J. Schaffner-Bielich,
Phys. Rev {\bf D 62}, 085008 (2000).

\bibitem{Mocsy:01prc}
O. Scavenius, A. Mocsy, I. N. Mishustin, D. H. Rischke,
Phys. Rev. {\bf C 64}, 045202 (2001).


\bibitem{Herpay:05}
T. Herpay, A. Patk\'{o}s, Zs. Sz\'{e}p and P. Sz\'{e}pfalusy,
Phys. Rev. {\bf D 71}, 125017 (2005).

\bibitem{Herpay:06}
T. Herpay and Zs. Sz\'{e}p, 
Phys. Rev. {\bf D 74}, 025008 (2006).

\bibitem{Herpay:07}
P. Kov\'acs and Zs. Sz\'{e}p,
Phys. Rev. {\bf D 75}, 025015 (2007).

\bibitem{Schaefer:07prd}
B. J. Schaefer and J. Wambach,
Phys. Rev. {\bf D 75}, 085015 (2007)

\bibitem{Schaefer:09}
B. J. Schaefer and M. Wagner,
Phys. Rev. {\bf D 79}, 014018 (2009).

\bibitem{Schaefer:08ax}
B.~J.~Schaefer and M.~Wagner,
Prog.Part.Nucl.Phys. {\bf 62}, 381 (2009)

\bibitem{Bowman:2008kc}
E.~S. Bowman and J.~I. Kapusta,
Phys. Rev. {\bf C 79}, 015202 (2009);
J.~I. Kapusta, and E.~S. Bowman,
Nucl.\ Phys.\  {\bf A 830}, 721C (2009).

\bibitem{koch}
L. Ferroni, V. Koch, and M. B. Pinto,
Phys. Rev. {\bf C 82}, 055205 (2010).

\bibitem{Andersen}
R. Khan and L. T. Kyllingstad, 
AIP Conf. Proc. {\bf 1343}, 504 (2011);
J. O. Andersen, R. Khan and L. T. Kyllingstad,
arXiv:1102.2779  

\bibitem{Fejos:8285}
G. Fejos, A. Patkos,
Phys. Rev. {\bf D 82}, 045011 (2010);
ibid {\bf D 85}, 117502 (2012).

\bibitem{Fejos}
G. Fejos, 
Phys. Rev. {\bf D 87}, 056006 (2013) ; arXiv:1212.3415

\bibitem{Schaefer:07}
B. J. Schaefer, J. M. Pawlowski and J. Wambach,
Phys. Rev. {\bf D 76}, 074023 (2007)

\bibitem{Braun}
J. Braun and H. Gies 
Phys. Lett. {\bf B 645} 53 (2007);
JHEP {\bf 06}, 024 (2006) 

\bibitem{gupta}
U. S. Gupta and V. K.Tiwari, 
Phys. Rev. D {\bf 81}, 054019 (2010).

\bibitem{Schaefer:09ax}
B. J. Schaefer, M. Wagner and J. Wambach,
Phys. Rev. {\bf D 81}, 074013 (2010)

\bibitem{Schaefer:09wspax}
B. J. Schaefer, M. Wagner and J. Wambach,
Proc. Sci., {\bf CPOD2009}, 017 (2009).

\bibitem{H.mao09}
H. Mao, J. Jin and M. Huang,
J. Phys. {\bf G 37}, 035001 (2010).


\bibitem{Herbst:2010rf}
T.~K. Herbst, J.~M. Pawlowski, and B.-J. Schaefer.
Phys. Lett. {\bf B 696}, 58 (2011). 

\bibitem{Pawl:Schaef}
B.-J. Schaefer, arXiv:1102:2772 [hep-ph];
J. M. Pawlowski, AIP Conf. Proc.{\bf 1343}, 75 (2011)

\bibitem{Marko:2010cd}
G.~Marko and Zs. Szep, 
Phys. Rev.{\bf D 82}, 065021 (2010).

\bibitem{kahara}
T. Kahara and K. Tuominen,
Phys. Rev. {\bf D 78}, 034015 (2008); ibid {\bf D 80}, 114022 (2009).
ibid {\bf D 82}, 114026 (2010).

\bibitem{Digal:01}
S. Digal, E. Laermann and H. Satz,
Eur. Phys. J. {\bf C 18}, 583 (2001). 

\bibitem{Ratti:06}
Claudia Ratti, Michael A. Thaler and Wolfram Weise,
Phys. Rev. {\bf D 73},  014019 (2006). 

\bibitem{Ratti:07}
S. R\"{o}{\ss}ner, C. Ratti,  and W. Weise,
Phys. Rev. {\bf D 75},  034007 (2007).

\bibitem{Hansen:07}
H. Hansen, W. M. Alberico, A. Beraudo, A. Molinari, M. Nardi and C. Ratti,
Phys. Rev. {\bf D 75}, 065004 (2007).

\bibitem{Ratti:07npa}
S. R\"{o}{\ss}ner, T. Hell, C. Ratti,  and W. Weise,
Nucl. Phys. {\bf A 814}, 118 (2008).

\bibitem{Tamal:06}
S. K. Ghosh, T. K. Mukherjee, M. G. Mustafa and R. Ray,
Phys. Rev. {\bf D 73}, 114007 (2006).

\bibitem{Sasaki:07}
C. Sasaki, B. Friman and K. Redlich,
Phys. Rev {\bf D 75}, 074013 (2007).

\bibitem{Hell:08}
T. Hell, S. R\"{o}{\ss}ner, M. Cristoforetti and W. Weise,
Phys. Rev {\bf D 79}, 014022 (2009).

\bibitem{Abuki:08}
H. Abuki, R. Anglani, R. Gatto, G. Nardulli and M. Ruggieri,
Phys. Rev {\bf D 78}, 034034 (2008).

\bibitem{Ciminale:07}
M. Ciminale, R. Gatto, N. D. Ippolito, G. Nardulli and M. Ruggieri,
Phys. Rev {\bf D 77}, 054023 (2008).

\bibitem{Fu:07}
W.-J. Fu, Z. Zhang and Y.-X. Liu,
Phys. Rev {\bf D 77}, 014006 (2008).

\bibitem{Kfuku:04plb}
K. Fukushima, 
Phys. Lett. {\bf B 591}, 277 (2004).

\bibitem{Fukushima:08d77}
K. Fukushima, 
Phys. Rev {\bf D 77}, 114028 (2008).

\bibitem{Fukushima:08d78}
K. Fukushima, 
Phys. Rev {\bf D 78}, 114019 (2008).

\bibitem{Fukushima:09}
K. Fukushima, 
Phys. Rev {\bf D 79}, 074015 (2009).

\bibitem{Contrera}
G. A. Contrera, M. Orsaria,and N. N. Scoccola,
Phys. Rev. {\bf D 82}, 054026 (2010). 

\bibitem{nonlocal}
A. E. Radzhabov, D. Blaschke, M. Buballa, and M. K. Volkov, 
Phys. Rev. {\bf D 83}, 116004 (2011)

\bibitem{Odilon}
O. Lourenco, M. Dutra, T. Frederico, A. Delfino, M. Malheiro
Phys.Rev. {\bf D 85}, 097504 (2012);
O. Lourenco, M. Dutra, A. Delfino, M. Malheiro
Phys.Rev. {\bf D 84}, 125034 (2011).

\bibitem{Ratti:07prd}
H. Hansen, W. M. Alberico, A. Beraudo, A Molinari, M. Nardi and C. Ratti
Phys. Rev. {\bf D 75},  065004 (2007).

\bibitem{Costa:04}
P. Costa, M. C. Ruivo, C. A. de Sousa and Yu. L. Kalinovsky
Phys. Rev. {\bf D 70},  116013 (2004).

\bibitem{Costa:05}
P. Costa, M. C. Ruivo, C. A. de Sousa and Yu. L. Kalinovsky
Phys. Rev. {\bf D 71},  116002 (2005).

\bibitem{Costa:09}
P. Costa, M. C. Ruivo, C. A. de Sousa, H. Hansen and W. M. Alberico
Phys. Rev. {\bf D 79},  116003 (2009).

\bibitem{Contrera:NLM}
G. A. Contrera, D. Gomez Dumm and Norberto N. Scoccola,
Phys. Rev. {\bf D 81}, 054005 (2010). 
\bibitem{Hiller}
A. A. Asipov, B. Hiller and Joao Da Providencia,
Phys.Lett. {\bf B 634}, 48-54 (2006);
A. A. Asipov, B. Hiller, A. H. Blin and Joao Da Providencia,
Annals Phys. (Amsterdam) {\bf 322}, 2021 (2007); 
A. A. Asipov, B. Hiller, J. Moreira, A. H. Blin and Joao Da Providencia,
Phys.Lett. {\bf B 646}, 91-94 (2007);
A. A. Asipov, B. Hiller, J. Moreira and A. H. Blin,
Phys.Lett. {\bf B 659}, 270-274 (2008);
B. Hiller,J. Moreira,A. A. Asipov and A. H. Blin,
Acta Phys.Polon.Supp. {\bf 5}, 1171-1177 (2012). 

\bibitem{tHooft:76prl}
G. 't Hooft, 
Phys. Rev. Lett. {\bf 37}, 8 (1976);
Phys. Rev {\bf D 14}, 3432 (1976).

\bibitem{fukushima:01}
K. Fukushima, K. Ohnishi, K. Ohta, 
Phys. Rev {\bf C 63}, 045203 (2001).

\bibitem{Skokov:2010sf}
V.~Skokov, B.~Friman, E.~Nakano, K.~Redlich, and B.-J. Schaefer,
Phys. Rev. {\bf D 82}, 034029 (2010)

\bibitem{Mizher:2010zb}
  A.~J.~Mizher, M.~N.~Chernodub and E.~S.~Fraga,
 Phys. Rev. {\bf D 82}, 105016 (2010).
\bibitem{Palhares:2008yq}
  L.~F.~Palhares and E.~S.~Fraga,
  Phys.\ Rev.\  D {\bf 78}, 025013 (2008)
  [arXiv:0803.0262 [hep-ph]].
  
\bibitem{Fraga:2009pi}
  E.~S.~Fraga, L.~F.~Palhares and M.~B.~Pinto,
  Phys.\ Rev.\  D {\bf 79}, 065026 (2009)
  [arXiv:0902.1498 [hep-ph]].
  
\bibitem{Palhares:2010be}
  L.~F.~Palhares and E.~S.~Fraga,
  Phys.\ Rev.\  D {\bf 82}, 125018 (2010)
  [arXiv:1006.2357 [hep-ph]].

\bibitem{Vivek:12}  
U. S. Gupta and V. K.Tiwari, 
Phys. Rev. D {\bf 85}, 014010 (2012).

\bibitem{Schaef:12}
B.-J. Schaefer and M. Wagner,
Phys. Rev. {\bf D 85}, 034027 (2012).

\bibitem{Sandeep}
Sandeep Chatterjee, Kirtimaan A. Mohan
Phys.Rev. {\bf D 85}, 074018 (2012);
ibid {\bf D 86}, 114021 (2012)

\bibitem{VKTR}
Vivek Kumar Tiwari, 
Phys. Rev. D {\bf 86}, 094032 (2012).
  
\bibitem{Weinberg:75}
S. Weinberg
Phys. Rev. {\bf D 11}, 3583 (1975).

\bibitem{Kapusta_Gale}
Finite Temperature Field Theory Principles and Applications, 
J. I. Kapusta and C. Gale,
Cambridge University Press.

\bibitem{Quiros:1999jp}
M.~Quiros, in Proceeding: The Summer School in High Energy Physics and Cosmology,
ICTP Series in Theoretical Physics, Trieste, Italy, 1998, edited by A. Masiero,
G. Senjanovic, and A. Smirnov (World Scientific , Singapore, 1999), Vol. 15, p. 436.
\bibitem{quench}
A. Bazavov and B. A. Berg
Phys. Rev. {\bf D 76}, 014502 (2007).
\end{thebibliography}
\end{document}